\documentclass[12pt]{article}
\pdfoutput=1
\usepackage{epsfig}
\usepackage{graphicx, color, xcolor}

\usepackage{amsmath}
\usepackage{amsfonts}
\usepackage{slashed}
\usepackage{amssymb}

\makeatletter

\@addtoreset{equation}{section}
\makeatother
\newcommand{\be}{\begin{equation}}
\newcommand{\ee}{\end{equation}}
\newcommand{\ben}{\begin{equation*}}
\newcommand{\een}{\end{equation*}}
\newcommand{\bea}{\begin{eqnarray}}
\newcommand{\eea}{\end{eqnarray}}
\newcommand{\alg}{\begin{align}}
\newcommand{\algx}{\end{align}}

\newcommand{\ppp}{\mkern-2mu}

\def\({\left(}
\def\){\right)}
\def\<{\left<}
\def\>{\right>}
\def\!{\right|}
\def\|{\left|}

\def\[{\left[}
\def\]{\right]}

\def\bq{{\bf q}}
\def\rmd{{\rm d}}
\def\E{{\cal{E}}}
\def\Ordo{{\cal{O}}}
\def\t{\tilde}

\def\mb{\mathbb}

\def\l{\ell}

\begin{document}

\begin{titlepage}
\vskip1cm
\begin{flushright}
\end{flushright}
\vskip1.55cm
\centerline{\Large \bf Conformal Janus on Euclidean Sphere}
\vskip1.5cm \centerline{ \large \textsc{Dongsu Bak$^{\, \tt a,e}$, Andreas Gustavsson$^{\, \tt b}$,  Soo-Jong Rey$^{\, \tt c,d,e}$
 }}
\vspace{1.5cm} 
\centerline{\sl  a) Physics Department,
University of Seoul, Seoul 02504 \rm KOREA}
 \vskip0.3cm
 \centerline{\sl b) School of Physics, Korea Institute for Advanced Study, Seoul 02455 \rm KOREA}
 \vskip0.3cm
 \centerline{\sl c) School of Physics \& Astronomy and Center for Theoretical Physics}
 \centerline{\sl Seoul National University, Seoul 08826 \rm KOREA}
 \vskip0.3cm
 \centerline{\sl d)
 Center for Theoretical Physics, College of Physical Sciences}
 \centerline{\sl 
 Sichuan University, Chengdu 610064 \rm PR CHINA}
 \vskip0.3cm
 \centerline{\sl e)
B.W. Lee Center for Fields, Gravity \& Strings}
 \centerline{\sl 
 Institute for Basic Sciences, Daejeon 34047 \rm KOREA}
\vskip0.3cm
 \centerline{\,\tt{dsbak@uos.ac.kr, agbrev@gmail.com, rey.soojong@gmail.com}\,} \vspace{1.5cm}

\centerline{ABSTRACT}
 \vspace{0.75cm} \noindent
We interpret Janus as an interface in a conformal field theory and study its properties. The Janus is created by an exactly marginal operator and we study its effect on the interface conformal field theory on the Janus. We do this by utilizing the AdS/CFT correspondence. We compute the interface free energy both from leading correction to the Euclidean action in the dual gravity description and from conformal perturbation theory  in the conformal field theory. We find that the two results agree each other and that the interface free energy scales precisely as expected from the conformal invariance of the Janus interface.
\vspace{1.75cm}
\end{titlepage}

\section{Introduction}
An interface refers to a $(d-1)$-dimensional subsystem (interface system) immersed inside a $d$-dimensional bulk system. It is known that critical behavior of the combined system is rich and highly nontrivial. The bulk system may be at a critical point or at off-critical point. At each cases, the interface system may separately be at critical or at off-critical point. As the parameters of bulk and interface systems are varied, the interface would undergo  variety of phase transitions. 

In this paper, we study a setup that both  bulk and interface systems are at criticality and that allow us to study its behavior via AdS/CFT correspondence.  
This is a typical situation that interactions in the bulk and in the boundaries are of the same order.  Equivalently, the interactions that drive the bulk into criticality also drive the interface into criticality. The critical behavior of bulk system is described by $d$-dimensional conformal field theory (CFT) and the critical behavior of interface system is described by $(d-1)$-dimensional conformal field theory. The total system is described by the interface CFT immersed inside the bulk CFT. 

\begin{figure}[ht!]
\vskip-1cm
\centering  
\includegraphics[width=13cm]{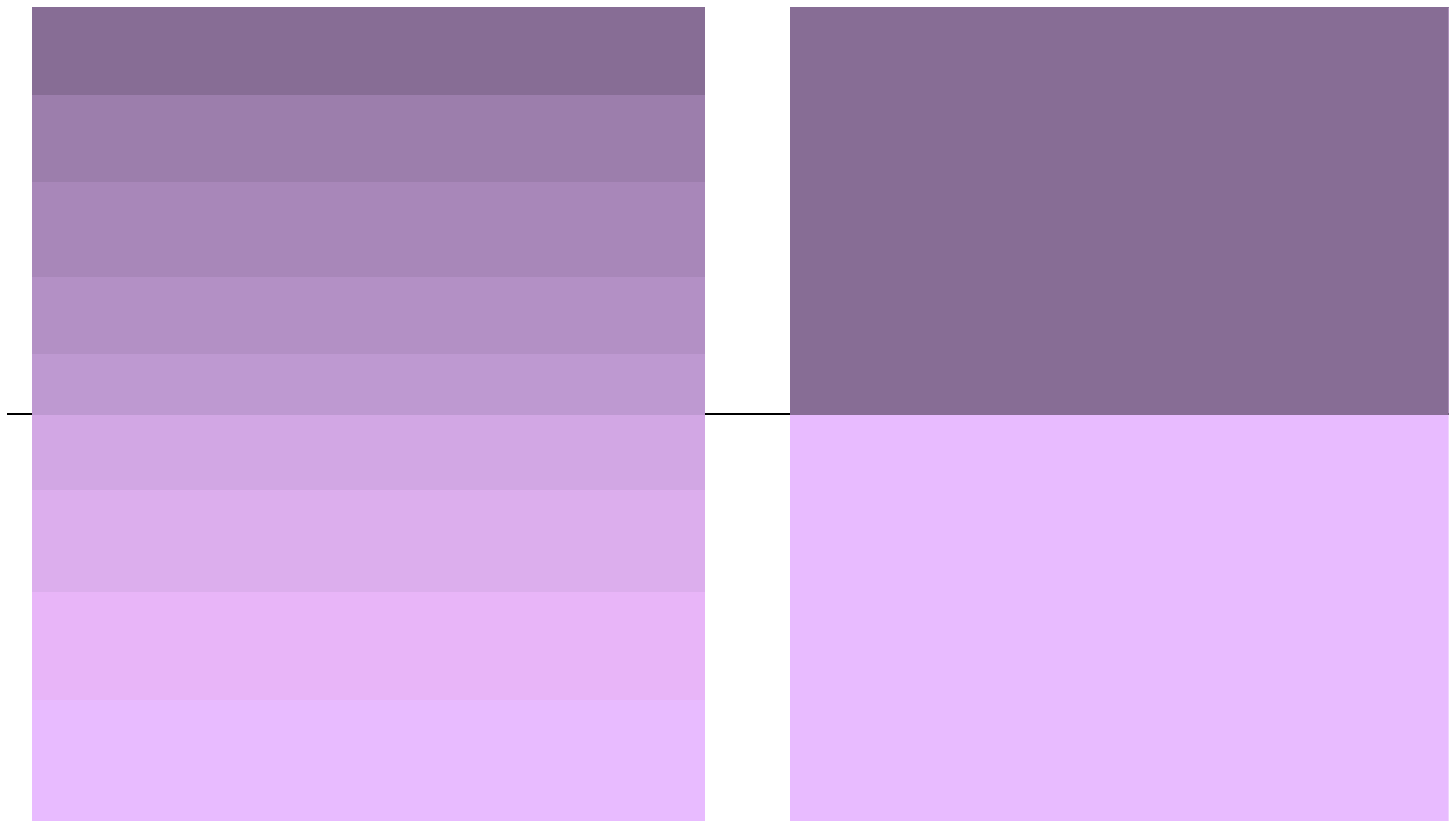}
\vskip-9.5cm
\caption{\sl An interface produced by a coupling parameter varying in vertical direction: (a) At off-criticality, the interface has a characteristic thickness. (b) At criticality, the interface is a codimension-one geometric surface.} 
\label{fig10}
\end{figure}

In the setup, the interface is given by a Janus deformation of bulk CFT, whose gravity dual is described by a Janus solution \cite{Bak:2003jk}. \footnote{For  recent discussions of 
Janus (related) systems, see  \cite{Papadimitriou:2004rz}-\cite{Bak:2013uaa} and references therein. There are also related studies of interface/domain-wall  partition function in supersymmetric gauge theories using localization  \cite{Hosomichi:2010vh}. }  This has the special feature that the Janus interface is constructible out of exactly marginal deformation of the bulk CFT, where the deformation parameter is interpreted as a varying coupling parameter of the exactly marginal operator. Across the interface, the coupling parameter interpolates from one asymptotic value to another, whose characteristic scale sets the thickness of the interface. In the regime the $d$-dimensional system is at criticality, this characteristic scale is driven to zero, so the deformation parameter jumps from one coupling constant to another 
across the interface. Moreover, given the argument a paragraph above, we expect that the system is described by a CFT$_d$ coupled to a CFT$_{d-1}$. This is somewhat surprising since one generally expects that the marginal deformation neither introduces any new degrees of freedom nor opens a new mass gap. 

In the setup above, the interface is a domain wall whose thickness is varied with the deformation of bulk CFT. In other words, Janus interface is a thick domain wall expanded around the thin wall limit. It should be noted, however, that there are also interfaces in conformal system whose thickness does not vary with deformation of the bulk CFT, i.e. intrinsically thin domain wall. In AdS/CFT setup, such interface can be engineered by D5-branes intersecting D3-branes with co-dimension one, providing a string theory setup for two-dimensional graphene interacting with strongly coupled three-dimensional conformal gauge system
\cite{Rey:2008zz}. There are also situations in which an interface CFT is realized inside non-conformal bulk in which the conformal symmetry is 
realized dynamically \cite{NakayamaRey}. 

An interesting point of the Janus deformation is that it is a hybrid of bulk CFT$_d$ and interface CFT$_{d-1}$, whose spacetime dimension differs by one. Systems in even and odd dimensions differ each other for their physical properties. 
Weyl and chiral symmetries are anomalous in even dimensions but intact in odd dimensions. Helmholtz free energy behaves very differently in even and odd dimensions. The Janus interface we study in this paper combines CFTs of even and odd dimensions. As such, one expects it furnishes a concrete setup for hybrid physical characteristics, exhibiting in one part even-dimensional behavior and in other part odd-dimensional behavior. We confirm such expectation from several one-point functions including renormalized free energy and stress tensor. We use the AdS/CFT correspondence and compute these physical observables, first from the gravity dual and then from dual CFT. With respect to the deformation parameter, we find complete agreement of both computations. 

This work is organized as follows. In section 3, we first Wick-rotate the system to Euclidean space and then put it on a $d$-dimensional sphere. We then turn on the exactly marginal coupling so that the Janus interface is located at the equator. We assume that the system admits large $N$ holography. In section 4, we construct the AdS$_{d+1}$ gravity dual, in which the Janus is obtained by exciting a scalar field. In sections 5, 6 and 7, using the gravity dual of section 4 and the AdS/CFT correspondence, we compute the free energy of the system. This physical observable diverges in the bulk infrared, and requires a regularization. By the AdS/CFT correspondence, in section 5, we relate this gravitational regularization to the regularization in the dual CFT. In sections 6 and  7, 
we extract the free energy of Janus interface for even and odd dimensions, respectively. We can also compute in section 8 the one-point function of the stress tensor. In section 9, we compute the interface free energy from the dual CFT by conformal perturbation theory.
In this section, we also construct some related Janus solutions whose boundary spacetimes are conformal to the $d$-sphere. 
  In section 10, we comment on the $g$-theorem
\cite{Affleck:1991tk} regarding renormalization group flow of the interface entropy. We argue that the interface free energy  is interpreted as the negative of the interface entropy. The change of multiple interfaces may either increase or decrease, depending on the signs of the deformation \cite{Bak:2013uaa} though the total entropy is always positive definite. 
We also relegate technical computations in the appendices.

\section{Summary of results} \label{sum}
We will let $\gamma$ parametrize the interface deformation, so that $\gamma=0$ corresponds to no interface. Then for a 2d CFT on $\mb{S}^2$ 
 with a Janus 1d interface on the equator, we get from the AdS side the following result for the partition function
\bea
Z(r,\epsilon,\gamma) &=& e^{-a_2\frac{r^2}{\epsilon^2}-a_1\frac{r}{\epsilon}-a\log \delta -F_I}
\eea
where $\epsilon$ is a cutoff near the boundary where the CFT lives in Fefferman-Graham coordinates and $\delta=\frac{\epsilon}{2r}$ as introduced 
below. The coefficients $a_2$ and $a$ are independent of $\gamma$, whereas $a_1$ and $F_I$ depend on $\gamma$. This reflects the fact that $a_2$ and $a$ have 2d origin, while $a_1$ and $F_I$ arise due to the 1d interface. As explained in 
\cite{Gaiotto:2014gha}, by forming the ratio
\bea
\frac{Z(r,\epsilon,\gamma)}{Z(r,\epsilon,0)} &=& e^{-b_1\frac{r}{\epsilon}-F_I}
\eea
where $b_1=a_1(\gamma)-a_1(0)$ and $F_I=F_I(\gamma)-F_I(0)$, we see that the divergences corresponding to $a_2$ and $a$ of the 2d CFT cancel. In particular the log divergences cancel out. We can then isolate the 1d interface theory contribution in a non-ambiguous way, and cancel the divergences corresponding to $b_1$ by adding a counter-term in the 1d interface theory, leaving us with a physical interface free energy $F_I$. We find that 
\bea
F_I(\gamma) &=& \frac{\ell}{4G}\log\sqrt{1-2\gamma^2}
\eea
where $\ell$ is the radius of AdS and $G$ is Newton's constant. In particular $F_I(0)=0$ as one would expect when there is no interface. We also notice 
that $F_I(\gamma)<0$ and that our interface Janus deformation breaks supersymmetry.

For 3d CFT on $\mb{S}^3$ with 2d Janus interface on the equator we get 
\bea
Z(r,\epsilon,\gamma) &=& e^{-a_3\frac{r^3}{\epsilon^3}-a_2\frac{r^2}{\epsilon^2}-a_1\frac{r}{\epsilon}-a_0\log\epsilon-F_I}
\eea
where $a_3$ and $a_1$ are constants independent of $\gamma$, while $a_2$, $a_0$ and $F_I$ depend on $\gamma$. Again we may isolate the 
contribution from the interface degrees of freedom by forming the ratio
\bea
\frac{Z(r,\epsilon,\gamma)}{Z(r,\epsilon,0)} &=& e^{-a_2\frac{r^2}{\epsilon^2}-a_0\log\epsilon-F_I}
\eea
In this case we may cancel the quadratic divergence and the log divergence by adding counterterms. The log divergence will then give rise to a conformal anomaly given by minus of the coefficient of the log divergence,
\bea
a_0 &=& \frac{\gamma^2 \ell^2 \pi}{16 G}+O(\gamma^4)
\eea
and the free energy is ambiguous. Still one should be able to, if one can identify the Fefferman-Graham cutoff with corresponding cutoff in the interface CFT, use this result to compare with an equally ambiguous field theory computation of the free energy, along similar lines as was done in four-dimensional super Yang-Mills theory in \cite{ReyZhou,Russo:2012ay}.

\section{Janus on Euclidean Sphere}
Begin with the bulk CFT$_d$ on $\mathbb{R}^d$ and the planar Janus interface  $\mathbb{R}^{d-1}$. By a conformal map, one can put the bulk CFT$_d$ on a $d$-dimensional sphere, $\mathbb{S}^d$ and an interface CFT$_{d-1}$ on the equator, which is a $(d-1)$-dimensional sphere.  Parametrize the $\mathbb{S}^d$ by 
\bea
\rmd s_d^2({\boldsymbol \Omega}) = r^2 (d \theta^2 + \sin^2 \theta \, \rmd s^2_{\mathbb{S}^{d-1}}({\boldsymbol \omega}) ), 
\label{sphere}
\eea 
where $r$ is the radius of the $\mathbb{S}^d$, $ \rmd s^2_{\mathbb{S}^{d-1}}$ is the metric of the $\mathbb{S}^{d-1}$ with unit radius. The bulk angular coordinates ${\boldsymbol \Omega}$ is split into the altitude angle $\theta$ ranging over $[0, {\pi}]$ and the interface angular coordinates ${\boldsymbol \omega}$. We then introduce an interface that halves the $\mathbb{S}^d$ at the equator, $\theta = \pi/2$. In general, this requires to introduce localized degrees of freedom on $\mathbb{S}^{d-1}$ that couple to the two sides $\mathbb{S}^d_\pm$.

\begin{figure}[ht!]
\centering  
\includegraphics[width=4.5cm]{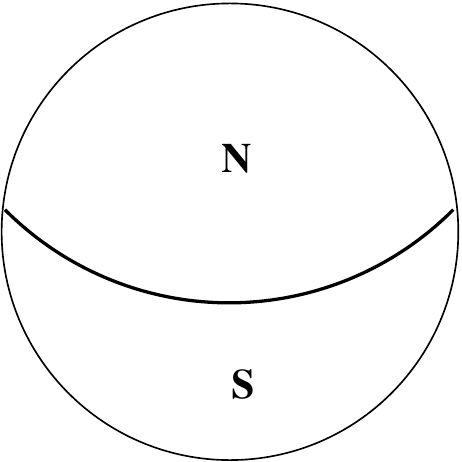}
\caption{\sl The Janus system on a sphere is depicted. On the northern/southern  hemisphere, we turn on an exactly marginal scalar operator with a coupling
$+\phi_I/-\phi_I$ respectively. 
}
\label{fig1}
\end{figure}

A distinguishing feature of the Janus interface is that it can be arranged from the bulk CFT$_d$ by simply turning on an exactly marginal scalar operator. Denote undeformed CFT$_d$ Lagrangian ${\cal L}_0({\boldsymbol \Omega})$ and its exactly marginal scalar operator ${\cal O}_\phi({\boldsymbol \Omega})$.  The deformed bulk CFT$_d$ is then defined by the action
\bea
I 
=\int_{\boldsymbol \Omega} {\cal L}_0 ({\boldsymbol \Omega}) + \int_{\boldsymbol \Omega} \lambda [\phi_B({\boldsymbol \Omega}) ]  
\, {\cal O}_{\phi} ({\boldsymbol \Omega}) \, . 
\eea
Here, in the spirit of AdS/CFT correspondence, we expressed the deformation parameter $\lambda$ 
in terms of boundary value of the bulk scalar field $\phi_B({\boldsymbol \Omega} )$ which is dual to the operator ${\cal O}_\phi$. The simplest and well-known Janus construction is when the operator ${\cal O}_\phi$ is given by the Lagrange density operator ${\cal L}_0$ of the bulk CFT$_d$. More generally, the Janus can be constructed with any scalar operators as long as they are exactly marginal. Indeed, in our considerations below, we only utilize the fact that the operator is exactly marginal scalar operator. 
The identification of precise functional form of the coupling parameter $\lambda(\phi_B)$ is a  complicated problem in a specific AdS/CFT correspondence. To the leading order in the bulk field expansion, one has in general 
\bea
\lambda(\phi_B({\boldsymbol \Omega}))= \phi_B (   {\boldsymbol \Omega}    ) + {\cal O}(\phi_B^2(   {\boldsymbol \Omega}   ) )\, . 
\label{couplingparameter}
\eea
The leading order is universal, while higher orders change with reparametrization of the bulk
 field. \footnote{For detailed discussion for the three-dimensional case, see Refs. \cite{Bak:2007jm, Chiodaroli:2009yw, Chiodaroli:2010ur}. 
 In this case, the bulk scalar $\phi$ describes the size modulus deformation of the target space.
The latter two references also include the discussion of half-BPS Janus system.}

For the Janus deformation with the interface at the equator, the gravity dual field $\phi_B ({\boldsymbol \Omega})$ takes the form
\bea
\phi_B(\theta) = \phi_I  \, \epsilon\Big({\frac{\pi}{2}-\theta}\Big) \, , 
\label{icoeff}
\eea 
where $\epsilon(x)$ is the sign function and $\phi_I$ is the deformation amplitude representing the jump of coupling parameter across the interface. The coupling parameter depends only on $\theta$, and so retains the stabilizer subgroup $SO(d)$ of the bulk $SO(d+1)$ isometry group.  The interface is distinguished by the topological quantum number, 
sign $\phi_I = \pm 1$. For the case of a single interface, without loss of generality, one can take the sign positive-definite.

\section{Gravity Dual}
It is known that the Janus interface geometry arises as a classical solution to the system of Einstein gravity coupled to negative cosmological constant and a minimal massless scalar field 
\be
I_{\rm gravity} = -\frac{1}{16\pi G}\int_{M_{d+1}} \left[ R -g^{ab} \partial_a \phi \partial_b \phi + \frac{d (d-1)}{\ell^2} \right] -{1 \over 8 \pi G} \oint_{ \partial M_{d+1}} K 
\label{einstein}
\ee
where the second term is  the Gibbons-Hawking boundary action \cite{Gibbons:1976ue}.
The boundary $\partial M_{d+1}$ on which the CFT$_d$ lives belongs to the conformal equivalence class of $\mathbb{S}^d$. 
The interface geometry can be found for arbitrary dimensions. For $(d+1) = 3$ and $5$, this system can be consistently embedded into the Type IIB supergravity and hence, via the AdS/CFT correspondence, microscopic understanding of dual interface CFT$_d$ system can be obtained \cite{Bak:2003jk, Bak:2007jm}. The scalar field here originates from the dilaton field of the underlying Type IIB supergavity and hence it is holographically dual to the CFT$_d$ Lagrangian density.  The equations of motion read
\bea
&& g^{ab} \nabla_a \partial_b \phi = 0 \nonumber \\
&& R_{ab} = -\frac{d}{\ell^2} g_{ab} +\partial_a \phi \partial_b \phi
\label{ein}
\eea 
The vacuum solution is the AdS$_{d+1}$ space with curvature radius $\ell$ and an everywhere constant scalar field. The Janus geometry is a nontrivial domain-wall solution in which the scalar field and metric approach those of the vacuum solutions. 

For instance, in three dimensions, $(d+1) = 3$, the Euclidean Janus geometry is given by \cite{Bak:2007jm}
\bea
&& \rmd s^2=\ell^2 \left[ \rmd y^2 + f(y) \, \rmd s^2_{M_2}\right]                        \cr
&& \phi(y)= \phi_0+ \frac{1}{\sqrt{2}} \log\left( 
\frac{1+\sqrt{1-2\gamma^2} +\sqrt{2} \gamma \tanh y}
{1+\sqrt{1-2\gamma^2} -\sqrt{2} \gamma \tanh y }
\right)
\label{3djanus}
\eea
where 
\bea
f(y)= \frac{1}{2}(1+\sqrt{1-2\gamma^2} \cosh 2y )
\eea 
and $0 \leq \gamma < \frac{1}{\sqrt{2}}$. The metric of 
$M_2$ has to satisfy 
\bea
R_{pq}(\bar{g})= - \bar{g}_{pq}
\eea
Below, we shall choose the metric of $ M_2$ as a global Euclidean $AdS_2$:  
\bea
\rmd s^2(M_2)= {1 \over \cos^2 \lambda} \left[ \rmd \lambda^2 + \sin^2 \lambda\,  \rmd \phi^2 \right]
\eea
where the fiber coordinate $\lambda$ ranges over $[0, \frac{\pi}{2}]$. 

\begin{figure}[ht!]
\centering  
\includegraphics[width=10cm]{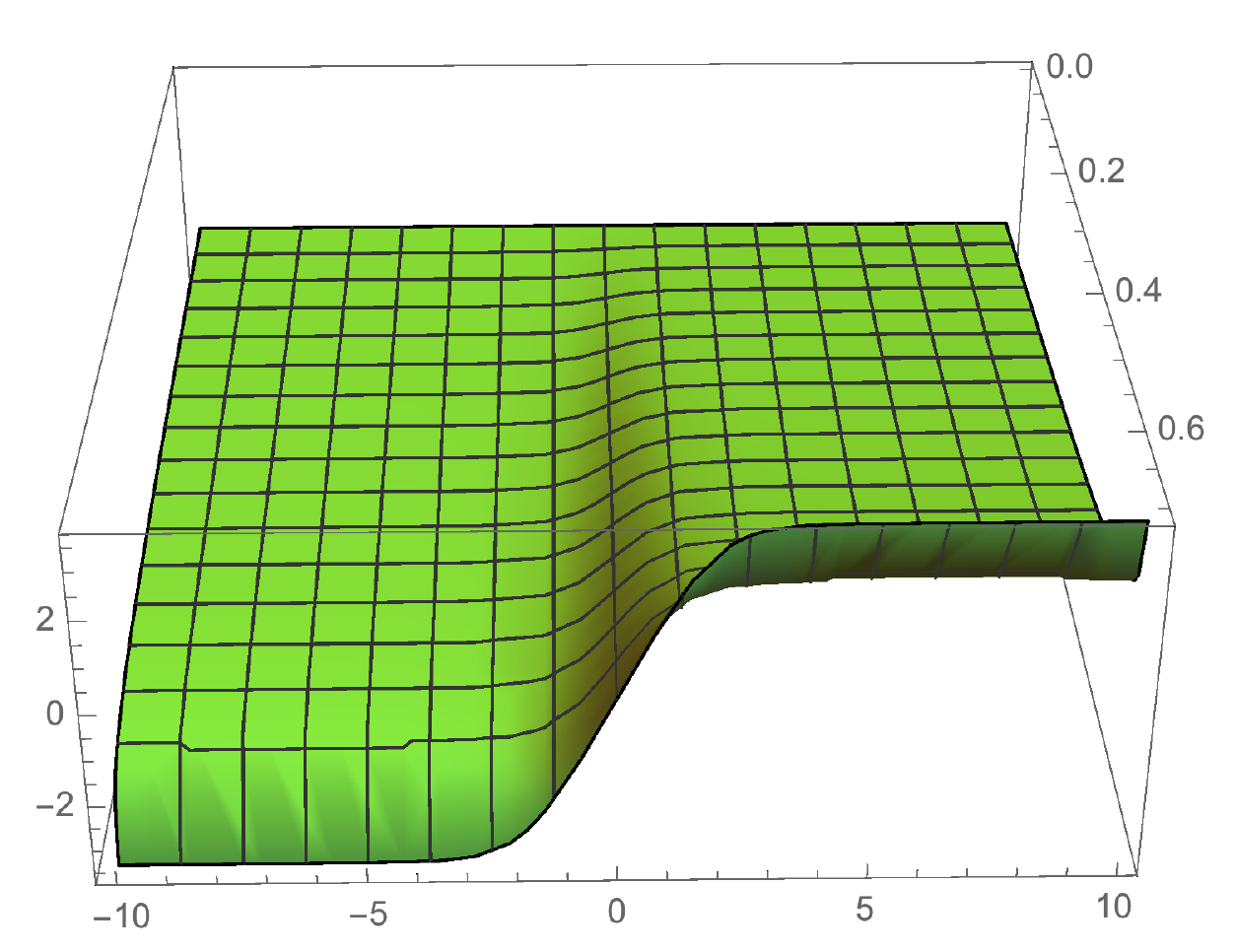}
\caption{\sl The Janus profile of minimal scalar field. The vertical axis covers $\phi(y) - \phi_0$, the horizontal axis covers $y = [-10, +10]$, and the depth covers entire domain of the deformation parameter $\gamma = [0, 1/\sqrt{2}=0.705...)$.   
}
\label{fig31}
\end{figure}

\begin{figure}[ht!]
\centering  
\includegraphics[width=10cm]{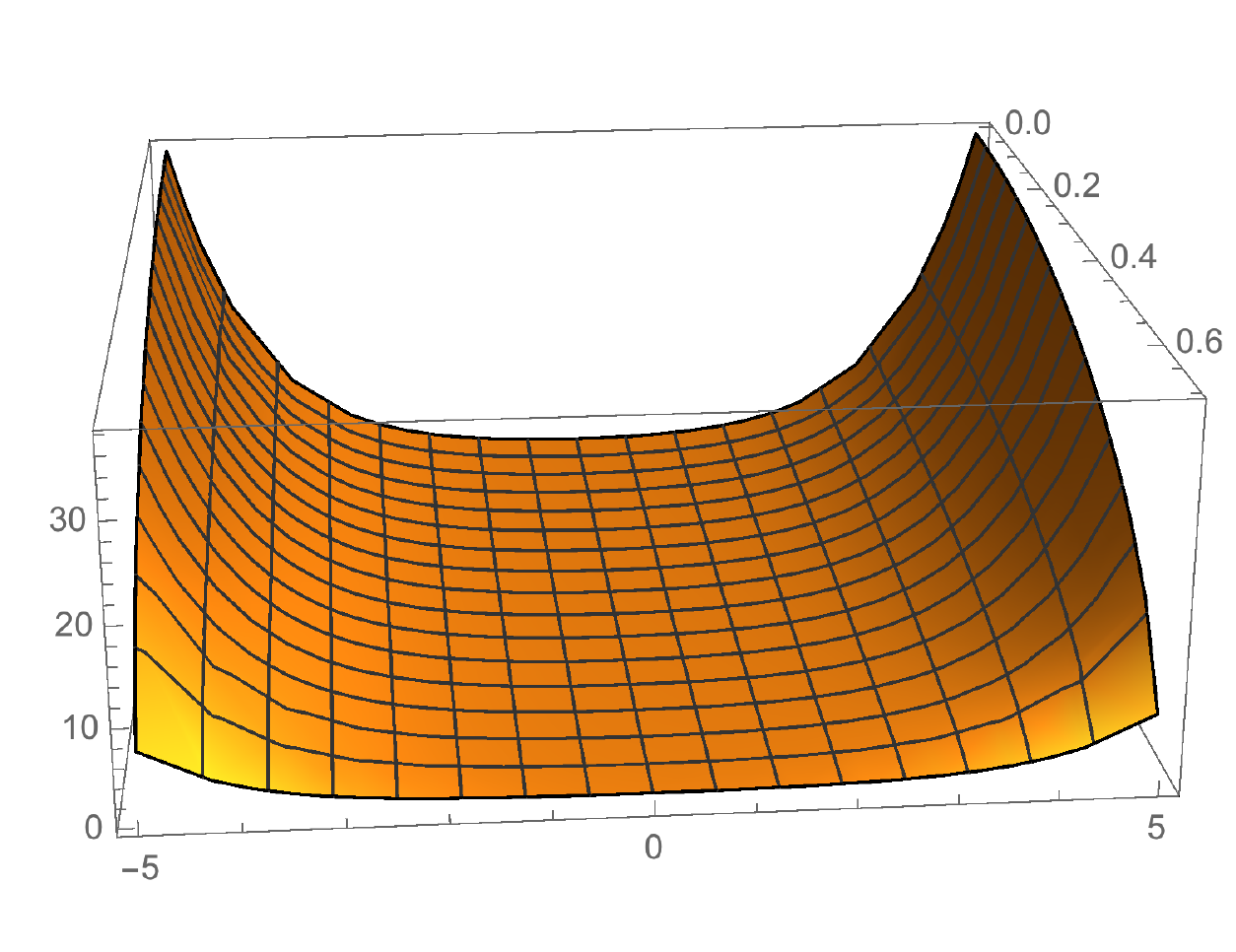}
\caption{\sl The Janus profile of metric function $f(y)$. The vertical axis covers $f(y)$, the horizontal axis covers $y = [-5, +5]$, and the depth encompasses entire domain of the deformation parameter $\gamma = [0, 1/\sqrt{2} = 0.705...)$.   
}
\label{fig32}
\end{figure}

The profile of the metric function $f(y)$ and the scalar field $\phi(y)$ are plotted over the entire range of the deformation 
parameter $\gamma$ in Figs. \ref{fig31} and \ref{fig32}.

The Janus geometry preserves the SO(2,1) isometry out of the SO(3,1) isometry, viz. Euclidean AdS$_2$ hypersurface inside Euclidean AdS$_3$ space. The conformal compactification of $y=\pm \infty$ is given by the boundary geometry of 
two hemispheres  $\mathbb{S}^2_\pm$, joined at the equator, $\lambda=\frac{\pi}{2}$.  
Hence, the entire boundary forms a full sphere $\mathbb{S}^2$, with an interface at the equator. 
We shall take the boundary metric as (\ref{sphere}). Without loss of generality, $\phi_0$ can be set to zero, so that the scalar field asymptotes to 
\bea 
 \phi(\pm \infty)=\pm  \frac{1}{\sqrt{2}} {\rm arctanh} \(\sqrt{2} \gamma\) 
 \label{scalarpro}
\eea
By the AdS/CFT correspondence, we identify these boundary values $\phi(\pm\infty)$ of the scalar field 
with $\pm \phi_I$. The scalar field is massless, so it sources an exactly marginal scalar operator in the dual CFT$_2$. 

More generally, in arbitrary dimensions, the gravity dual of an exactly marginal deformation is again
described by  
the action (\ref{einstein}). The
Euclidean Janus geometry is given by two patches, labeled by $\pm$ \cite{Bak:2015jxd},
\bea
\cr
\rmd s_\pm^2\ &=& \frac{\ell^2}{q^2_\pm} \left[\frac{ \rmd q^2_\pm}{P(q_\pm)} +   \rmd s^2 _{M_{d}}\right]     \cr
\cr
\cr
\phi_{\pm}(q_\pm)&=& \phi_0  \pm \gamma \int_{q_\pm}^{q_*} \rmd x  \frac{ x^{d-1}}{\sqrt{P(x)}}, 
\label{d+2janus}
\eea 
where $P(x)$ is dimension-specific polynomial
\bea
P(x)=1-x^2 + \frac{\gamma^2}{d(d-1)} x^{2d}
\eea 
and $q_*$ denotes the smallest positive root of $P(q_*)=0$. Here, $q_\pm$ are parameters
ranging over $[0, q_*]$ and
$\gamma$ is the deformation parameter ranged over $[\, 0,\sqrt{d-1}\left(\frac{d-1}{d}\right)^{\frac{d-1}{2}}\,)$.  The metric of the hypersurface $M_d$ has to satisfy the hyperbolicity 
\bea
R_{pq}(\bar{g})= - (d-1)\bar{g}_{pq} \,.
\eea
We choose the metric of $M_d$ as
\bea
\rmd s^2 (M_d)= \frac{1}{\cos^2 \lambda} [\rmd \lambda^2 + \sin^2 \lambda \, \rmd s^2(\mathbb{S}^{d-1})] \,.
\eea

To cover the entire space, we need to choose the coordinates $q_{\pm}$ of the two respective patches ranged over the same interval. To see this clearer, we revisit the three-dimensional solution, $(d+1)=3$, 
and rename the $y$ coordinate as 
\bea
q_+ &=& \frac{1}{\sqrt{f(y)}}\qquad \qquad (y\in[0,+\infty])\cr
q_- &=& \frac{1}{\sqrt{f(y)}} \qquad \qquad (y\in [-\infty,0]) \,.
\eea
We see that $q_\pm\rightarrow 0$ correspond to $y\rightarrow \pm \infty$ and $q_{\pm} = q_*$ corresponds  to $y=0$. The entire domain of $y = (-\infty, + \infty)$ is covered by two identical copies of $q_\pm = [0, q^*)$. For general dimension $d$, provided the scalar field is differentiable, the two patches are smoothly joined at $q_{\pm} = q_*$. In the asymptotic regions, $q_\pm \rightarrow 0$, the scalar field takes the asymptotic values $\phi(\pm\infty) = \phi_0 \pm \gamma \int_{0}^{q_*} dg  \frac{ g^{d-1}}{\sqrt{P(g)}}$ where the sign depends on the respective coordinate patch used. The Janus geometry has SO($d$,1) isometry of the AdS$_d$ hypersurface out of the SO($d+1$,1) isometry of the Euclidean AdS$_{d+1}$ space. By the AdS/CFT correspondence, the holographic dual CFT$_d$ is deformed by an interface that preserves $(d-1)$-dimensional conformal invariance. We shall refer to the latter system as interface CFT (ICFT). In the gravity dual, the minimal scalar field is massless, so it couples to an exactly marginal scalar operator in dual CFT$_d$. We have just shown that the Janus geometry provides an elegant construction of interface while preserving the conformal invariance both at the bulk and the interface.    


\section{Renormalized Free Energy}
To understand the ICFT better, we now compute physical observables. The simplest observable is the free energy, the expectation value of an identity operator. In this section, we compute the free energy of the ICFT by computing the classical, on-shell Euclidean action of the gravity dual. The classical  Euclidean action is infrared divergent, so we shall be computing it using the method of holographic renormalization \cite{deHaro:2000vlm}. In this method, the first step is to regularize the action of gravity dual by introducing an infrared cut-off at timelike infinity, where the geometry asymptotes to AdS$_{d+1}$. For the vacuum solution (in which the Janus deformation parameter is put to zero), the cutoff will be chosen such that it retains the stability isometry subgroup maximal, namely, the induced boundary metric is a $d$-dimensional sphere, $\mathbb{S}^d$. The second step is to add counter-terms to cancel divergences as the infrared cutoff is removed. Such a cutoff can be chosen in any coordinate system one adopts and different choices correspond to different subtraction schemes.  
In general, these schemes differ from one another by the amount of finite subtractions in addition to the infrared divergences. Among them, the minimal subtraction scheme, viz. the scheme that only subtracts the infrared divergence, is provided by the   
Fefferman-Graham (FG) coordinates. 

Therefore, as a preliminary step, we first exercise out the computation of renormalized free energy for the undeformed (without Janus interface) CFT$_d$ on $\mathbb{S}^d$, described both in FG coordinate system and in other coordinate system. From the computation, we explicitly find that we obtain the finite subtraction different in the two coordinate systems. We shall hence adopt the FG coordinate system in this section 
and extract the free energy in the minimal subtraction scheme. 

\subsection{Free Energy in Fefferman-Graham Coordinates}\label{sdfreeenergy}
Start with the undeformed CFT$_d$ on $\mathbb{S}^d$ of radius $r$. Its gravity dual is described by
the Euclidean AdS$_{d+1}$ space
\bea
ds^2=\ell^2 \left[ d\rho^2 + \sinh^2 \rho \, ds^2 ({\mathbb{S}^d}) \right]
\label{dualsphere}
\eea
It turns out that the FG coordinate has a complication for the evaluation of the free energy.

We compute the free energy of the undeformed CFT$_d$ on $\mathbb{S}^d$ in the FG scheme, viz. by 
regularizing infrared divergences in the FG coordinate system. In general, one can always put the FG coordinate system in the form
\bea
\rmd s^2 := g_{ab} \rmd x^a \rmd x^b = \ell^2 \left[
\frac{\rmd u^2}{u^2} +\frac{1}{u^2}\, h_{ij}(x, u^2)\, \rmd x^i \rmd x^j 
\right] \qquad\qquad (\, 0 \le u  \, 
). 
\label{fgmetric}
\eea
The on-shell action of the gravity dual diverges in the infrared. We regularize it in the FG scheme by cutting off the FG geometry at $u=\epsilon$ with $u \ge \epsilon$.
Taking into account  the Gibbons-Hawking boundary action at the cut-off, 
the full regularized action reads
\bea
I_{\rm reg}=-\frac{1}{16\pi G}\int_{M_\epsilon} \rmd^{d+1} x \, \sqrt{g} \left[ R -g^{ab} \partial_a \phi \partial_b \phi + \frac{d (d-1)}{\ell^2} \right]
-\frac{1}{8\pi G}\int_{\partial M_\epsilon} \rmd^{d} x \, \sqrt{\gamma} K
\eea
where $\gamma_{ij}$ is the induced metric at the surface $u=\epsilon$. $K$ is the trace of the extrinsic curvature $K_{ab}$. Recall that, with the surface normal unit vector $n$ specified by
$n_u= -\frac{\ell}{u}, n_i=0$, the extrinsic curvature $K_{ab}$ is defined by
\bea
K_{ab}=\frac{1}{2} {\cal L}_n (g_{ab} -n_a n_b)\,,
\eea
where ${\cal L}_n$ is the Lie derivative along $n$.
Using the Einstein field equation (\ref{ein}),  the regularized on-shell action becomes
\bea
I_{\rm reg}=I_{\rm bulk}+I_{\rm suface} \, , 
\eea
where
\bea
&& I_{\rm bulk} =\frac{d}{8\pi G\, \ell^2}\int_{M_\epsilon} \rmd^{d+1} x \, \sqrt{g} \cr
&& I_{\rm suface}=-\frac{1}{8\pi G}\int_{\partial M_\epsilon} \rmd^{d} x \, \sqrt{\gamma}\, K \, . 
\eea
In the above FG coordinates system, we find that 
\bea
&& I_{\rm bulk}=\frac{d \ell^{d-1}}{8\pi G}\int \rmd^{d} x \int_{\epsilon} \frac{\rmd u}{u^{d+1}} \sqrt{h} \cr
&& I_{\rm suface}=-\frac{\ell^{d-1}}{8\pi G \, \epsilon^{d}}\int \rmd^{d} x \left(
1-\frac{1}{d} u \partial_u
\right)\sqrt{h}\, \Big|_{u=\epsilon}  \, . 
\eea
The regularized action $I_{\rm reg}$ in general has an expansion \footnote{With Janus deformation below, the singular  terms of remaining powers, $b_{(1)}\epsilon^{-d+1}+ b_{(3)}\epsilon^{-d+3}+ \cdots $  
may appear and those should be subtracted in addition  by counter-terms.} 
\bea
I_{\rm reg}= \frac{\ell^{d-1}}{16\pi G} \int \rmd^d x \, \sqrt{h_{(0)}}\left( 
\frac{a_{(0)}}{\epsilon^d}+\frac{a_{(2)}}{\epsilon^{d-2}}+\cdots  -2 \log(\epsilon) a_{(d)}
\right) +O(\epsilon^0), 
\eea
where the logarithmic contribution exists only when $d$ is even. In the holographic renormalization, one chooses the counter-term as
\bea
 I_{\rm ct}=- \frac{\ell^{d-1}}{16\pi G} \int \rmd^d x \, \sqrt{h_{(0)}}\left( 
\frac{a_{(0)}}{\epsilon^d}+\frac{a_{(2)}}{\epsilon^{d-2}}+\cdots  -2 \log(\epsilon) a_{(d)}
\right)
\eea 
such that in the limit $\epsilon \rightarrow 0$ of the renormalized action $I_{\rm ren}= I_{\rm bulk}+ I_{\rm surface} + I_{\rm ct}$ 
all the singular divergences are subtracted, while leaving the finite contribution intact.

We now apply the above general consideration to the metric (\ref{dualsphere}). The FG coordinates are identified to be 
\bea
u=2 r e^{-\rho}
\eea
and
\bea
h_{ij} \, \rmd x^i \rmd x^j =    \left(1-\frac{u^2}{4 r^2}\right)^2 r^2 \, \rmd s^2 (\mathbb{S}^d) \, . 
\eea 
The regularized action takes the form
\bea
I_{\rm reg}=  \frac{\mbox{Vol}(\mathbb{S}^d)}{16\pi G} d \, (\ell/2)^{d-1} \left(
A_d + B_d
\right)
\eea
where $A_d$ and $B_d$ are contributions from $I_{\rm bulk}$ and $I_{\rm surface}$, respectively. For the above metric, they take the forms
\bea
&& A_d = \int^1_{\delta} \rmd z \frac{1}{z^{d+1}} (1-z^2)^d \cr
&& B_d = -\frac{1}{\delta^d} \left(
1-\frac{1}{d} z \partial_z
\right)(1-z^2)^d \,\Big|_{z=\delta} \, , 
\eea
where the parameter $\delta$ is related to the cut-off $\epsilon$ in the FG coordinate $u$ by 
\bea
\delta = \frac{\epsilon}{2r}.
\label{cutoff0}
\eea 
It is illuminating to work out explicitly for lower dimensions. 

For $d=2$, one finds
\bea
&& A_2= \frac{1}{2} \left(\frac{1}{\delta^2} -\delta^{2} \right) + 2 \log \delta \cr
&& B_2 =-  \left(\frac{1}{\delta^2} -\delta^{2} \right) \, . 
\eea
The renormalized action reads \footnote{For $d$ even, the renormalized 
action includes regularization-dependent, non-universal contribution. To resolve any ambiguity in the correspondence, one has to specify 
how the regularization is done from the view point of  the both sides. Here, our choice 
is in such a way that the regularization is independent of the couplings. See \cite{ReyZhou, Russo:2012ay,  
Huang:2014gca, Gomis:2015yaa} for discussions in this context.}
\bea
I_{\rm ren} = -\frac{c}{3} \log (2 r) \, . 
\eea
Here, we identified the central charge of the CFT$_d$ with the Brown-Henneaux \cite{Brown:1986nw} or Henneaux-Rey \cite{Henneaux:2010xg} central charge $c = \frac{3\l}{2G}$. This identification also agrees with the central charge derived by the AdS/CFT correspondence from the Weyl anomaly 
\cite{Henningson:1998gx}. We can extract the Weyl anomaly integrated over the boundary sphere from the coefficient of the $\log \epsilon$ term in the regularized action above. \footnote{To compare our result with \cite{Henningson:1998gx}, one should note the different conventions. First we use a FG coordinate $u$ that has a double pole, while  they use a FG coordinate that has a simple pole at the boundary. This accounts for a factor of $2$ in the definition of the Weyl anomaly. Second, they use a convention where Riemann tensor has opposite sign compared to us. By noticing this, we find that our result can be reproduced from their more general result by specializing to a round two-sphere boundary with curvature scalar $R=2/r^2$ in our convention.} Adding the renormalization point scale $\mu$ appropriately, we finally obtain
 \bea
I_{\rm ren} = -\frac{c}{3} \log (r \mu). 
\label{d2freeenergy}
\eea
Hence, the renormalized partition function is obtained as 
\bea
Z_{\rm ren}:= \exp ( - I_{\rm ren}) = (r \mu)^{\frac{c}{3}} \, . 
\eea

For $d=3$, one finds 
\bea
&& A_3=\frac{16}{3} +  \frac{1}{3 \delta^3} -\frac{3}{ \delta}-3\delta +\frac{\delta^3}{3} \cr
&& B_3 =-  \frac{1}{ \delta^3} +\frac{1}{ \delta}+\delta -{\delta^3} \,.
\eea
Therefore, the renormalized action reads \cite{Marino:2011nm}
\bea
I_{\rm ren} =\frac{\pi \ell^2}{2 G}. 
\eea
The renormalized partition function
\bea
Z_{\rm ren} = \exp \left( - {\pi \ell^2 \over 2 G} \right)
\eea
is independent of scale and hence ambiguity-free.

For $d=4$, one finds
\bea
&& A_4=  \frac{1}{4 \delta^4} -\frac{2}{ \delta^2}+2\delta^2 -\frac{\delta^4}{4}
 -6 \log \delta
 \cr
&& B_4 = -\frac{1}{ \delta^4} +\frac{2}{ \delta^2}-2\delta^2 + {\delta^4} \, . 
\eea
The renormalized action reads
\bea
I_{\rm ren} =\frac{ \pi \ell^3}{2 G} \log (2r) \, . 
\eea
The result fits perfectly with the Weyl a-anomaly of four-dimensional ${\cal N}=4$ $SU(N)$ gauge theories  \cite{Pestun:2007rz} and ${\cal N}=2$ $(SU(N))^{\otimes_k} $ quiver gauge theories \cite{Rey:2010ry}
on $\mathbb{S}^4$
\bea
a= \frac{\pi \ell^3}{2 G} = a_o N^2 \, , 
\eea
with $a_o=1$ and $a_o = k$, respectively. 
Reinstating the renormalization point scale $\mu$, the renormalized partition function reads
\bea
Z_{\rm ren} = (\mu \, r)^{-a_o N^2}.
\eea
%

\subsection{Free Energy in Other Coordinates}
We now compute the renormalized free energy of the undeformed CFT$_d$ in other scheme, viz. in other coordinate system. The metric in (\ref{dualsphere}) can be written as
\bea
\rmd s^2 = \ell^2 \left[ \rmd y^2 + \frac{\cosh^2 y}{\cos^2 \lambda}  \left( \rmd \lambda^2 + \sin^2 \lambda\,  \rmd s (\mathbb{S}^{d-1})^2  \right) \right] \, . 
\label{othercoor}
\eea
We shall cut off at  the infrared along the hypersurface 
\bea
r \frac{\cos \lambda}{\cosh y} =\epsilon_1
\label{cutoff11}
\eea
This choice of the cutoff turns out to agree with the FG cutoff described in the subsection \ref{sdfreeenergy} provided the cutoff $\epsilon_1$ here is identified with an appropriate function of the FG cutoff $\delta$. To see this explicitly, we note that the coordinates in  (\ref{dualsphere})
are related to the ones in (\ref{othercoor}) by
\bea
  \frac{\cos \lambda}{\cosh y} &=&\frac{1}{\cosh \rho} \qquad \mbox{and} \qquad
 \sin\lambda = \tanh \rho \sin \theta \, . 
\eea
Thus, it is clear that the cutoff hypersurface (\ref{cutoff11}) describes the same hypersurface as constant $\rho$ in the coordinate system (\ref{dualsphere}). The precise relation between $\epsilon_1$ and $\delta$ will be relegated to the next subsection.

For the explicit computation, let us focus on the three-dimensions, $d=3$. With the cutoff (\ref{cutoff11}), 
the bulk action becomes
\bea
I_{\rm bulk} = \frac{1}{4\pi G\, \ell^2}\int_{M_\epsilon} \rmd^{3} x \, \sqrt{g}  = \frac{\ell}{G}\int ^{y_0}_0 \rmd
y \, {\cosh^2 y} 
\int^{\lambda(y)}_0 \frac{\rmd \lambda |\sin \lambda|}{\cos^2 \lambda} \, , 
\eea
where $\cosh y_0 = \frac{r}{\epsilon_1}$ and $\cos \lambda(y) =\frac{\epsilon_1}{r} \cosh y $.
The result is
\bea
I_{\rm bulk} = \frac{\ell}{2 G} \left[ \frac{r^2}{\epsilon_1^2} \sqrt{1-\frac{\epsilon_1^2}{r^2}} +
\log\left(\frac {\frac{\epsilon_1}{r}} {1+ \sqrt{1-\frac{\epsilon_1^2}{r^2}}}\right) \right] \, . 
\eea
Suppose we identify
\bea
\frac{\frac{\epsilon_1}{r}}{1+ \sqrt{1-\frac{\epsilon_1^2}{r^2}}} =\delta. 
\label{eps0eps1}
\eea
Then, one finds that the bulk action 
\bea
I_{\rm bulk} = \frac{\ell}{2 G}\left[ \frac{1}{4} \left(\frac{1}{\delta^2} -\delta^{2} \right) + \log \delta \right]
\eea
agrees with the bulk action in the FG coordinate system. In the present coordinate system, one also finds that $I_{\rm surface}$ is given by
\bea
 I_{\rm surface} = -  \frac{\ell}{G}  \left( \frac{r^2}{\epsilon_1^2} \sqrt{1-\frac{\epsilon_1^2}{r^2}} \right)
 =-  \frac{\ell}{4 G}  \left(\frac{1}{\delta^2} -\delta^{2} \right). 
\eea
This again agrees with the surface action contribution computed in the FG coordinate system.  
We relegate details of the computation to  appendix \ref{appb}. 

We draw the conclusion that, while different coordinate system gives in general different subtraction schemes,  appropriate relation between cutoffs can be specified to prescribe identical subtraction scheme and hence the same renormalized free energy.  
This prompts us to understand precise relation between infrared cutoffs in the holographic renormalization, to which we now turn in the next subsection.  

\subsection{Relations between Cutoffs}
In the holographic renormalization, the FG scheme is considered the most convenient as it subtracts power divergences only. The lesson of the last subsection was that the FG scheme can be made not only in FG coordinate system but also in any other coordinate systems {\sl provided} each respective cutoff is correspondingly related to 
each other. Below, we find explicit relations between cutoffs  in different coordinate system that all lead to the minimal subtraction.
 
We first define the cutoff $\epsilon$ of FG minimal subtraction scheme by the following hypersurface in the  global AdS coordinate system (\ref{dualsphere}) and the FG coordinate system (\ref{fgmetric}) 
\bea
2r e^{-\rho_{\infty}} &=& \epsilon \cr
u_0 &=& \epsilon \, . 
\label{cutoff1}
\eea
We also define the cutoff $\epsilon_1$ in the other coordinate system (\ref{othercoor})
\bea
\frac{r}{\cosh \rho_{\infty}} &=& \epsilon_1 \, . 
\label{cutoff2}
\eea
As explained in the last subsection, this cutoff is not independent but leads to the same cutoff as the FG scheme. As such, we used the cutoff position $\rho_{\infty}$ the same value for either choices of the cutoff. 

To relate the two cutoffs, we find it convenient to introduce dimensionless cutoff parameters by (as was done for the first in (\ref{cutoff0}))
\bea
\delta &=& \frac{\epsilon}{2r}\cr
\cr
\delta_1 &=& \frac{\epsilon_1}{r} \, . 
\eea
From (\ref{cutoff1}) and (\ref{cutoff2}), one finds the relation between the two cutoff parameters as
\bea
\frac{1}{\delta} + \delta &=& \frac{2}{\delta_1} \, . \label{cutoffs}
\eea
This can be inverted. We find the desired relation for the FG minimal subtraction scheme as
\bea
\delta &=& \frac{\delta_1}{1+\sqrt{1-\delta_1^2}} . 
\eea

\section{Free Energy of   Janus CFT$_{2}$}
\label{icft2}

Having understood schemes for holographic renormalization, we now extract the renormalized free energy of the ICFT$_d$. In this section, via the AdS/CFT correspondence, we shall first compute the free energy from the gravity dual. We shall focus on the three-dimensional gravity dual, the Janus geometry (\ref{3djanus}).

The subtraction scheme and renormalization thereof must preserve all symmetries the system retains (apart from the Weyl anomaly for $d$ even and nontrivial curvature background). The infrared cutoff needs to be chosen accordingly. For the gravity dual of undeformed CFT$_d$, we saw in the previous section that the FG scheme works perfectly, since it 
simply corresponds to a foliation of the Euclidean AdS$_{d+1}$ space by $\mathbb{S}^d$ hypersurfaces,  
for which the SO($d+1$) isometries are manifest. For the Janus interface, however, these SO($d+1$) isometries are broken to SO($d$). Accordingly, the cutoff hypersurfaces must be chosen such that the SO($d+1$)/SO($d$) coset is nonlinearly realized. 
Indeed, for the Janus geometry dual to the ICFT$_d$ on $\mathbb{R}^{1, d-1}$, the requisite FG coordinate system was constructed in \cite{Papadimitriou:2004rz}. There, it was pointed out that the FG coordinate system does not cover the entire bulk region: the  wedge-shaped bulk region emanating from the boundary location of the interface is not covered. This implies that the FG coordinate system constructed in \cite{Papadimitriou:2004rz} is not globally well-defined and it has to be further analytically extended. 

In this work, we do not attempt to construct a FG  coordinate system and its analytic extension thereof. Instead,  we introduce a coordinate $v$  by
\bea
v=r  \frac{\cos \lambda}{\sqrt{f(y)}}
\label{newrelation}
\eea
and simply declare our cutoff for the minimal subtraction scheme by the hypersurface $v=\epsilon_1$. We are motivated to adopt this scheme since this choice simply replaces the $\cosh y$ factor in the undeformed geometry in (\ref{othercoor}) by the square-root of the scale function $f(y)$ in the Janus geometry (\ref{3djanus}). 
We can further justify this choice by the following observation.  At a short distance away from the interface, the corresponding bulk geometry takes near the cutoff hypersurface the same form as the undeformed one once we ignore the higher-order terms that do not contribute to the renormalized action. Later, we will show that this observation holds for arbitrary dimensions. 

Thus, for the computations below, we shall adopt the coordinate system $(v,\lambda,{\phi})_+ \oplus (v,\lambda,{\phi})_-$ by eliminating the coordinate $y$ using the above relation (\ref{newrelation}). This coordinate system consists of two branches, $\pm$, coming from the region with positive/negative 
$y$, respectively. 
  
We find that 
\bea
I_{\rm bulk}= I^0_{\rm bulk}+\Delta I_{\rm bulk} \, , 
\eea
where
\bea
 I^0_{\rm bulk} &=& \frac{\ell}{2 G}\left[ \frac{1}{4} \frac{1}{\delta^2}  + \log \delta + \Ordo(\delta)\right]\cr
 \Delta I_{\rm bulk}(\gamma) &=&  \frac{\ell}{2 G}\left[ \frac{1}{\delta} \alpha(\sqrt{1-2\gamma^2}) -\frac{1}{2}\log  \frac{1}{\sqrt{1-2\gamma^2}}
 + \Ordo(\delta)
 \right] \, . 
\eea
We introduce the function $\alpha(z)$ by
\bea
\alpha(z)&=&\frac{1-z}{ \sqrt{z}} \int_0^1 \frac{dx}{1+x^2 +\sqrt{1+x^4 +\frac{2}{z} x^2}} \cr
&=& \frac{\sqrt{1+z}}{\sqrt{2}} \left[
{\bf K}\Big(\frac{1-z}{1+z}\Big) - {\bf E}\Big(\frac{1-z}{1+z}\Big)\right]\, , 
\eea
where ${\bf K}(k^2)$ and ${\bf E}(k^2)$ are, respectively,  the first kind and the second kind of the complete elliptic integral defined by
\bea
{\bf K}(k^2) &=& \int^1_0 dx \frac{1}{\sqrt{1-x^2}\sqrt{1-k^2 x^2}}\cr
{\bf E}(k^2) &=& \int^1_0 dx \frac{\sqrt{1-k^2 x^2}}{\sqrt{1-x^2}} \, . 
\eea
For small $\gamma$, the function $\alpha(\sqrt{1-2\gamma^2})$ is expanded as
\bea
\alpha(\sqrt{1-2\gamma^2})= \frac{\pi}{8} \gamma^2 + \frac{15\pi}{128} \gamma^4+ \frac{315\pi}{2048} \gamma^6+\Ordo(\gamma^8) \, . 
\eea
For the surface action, we also obtain
\bea
I_{\rm surface}= I^0_{\rm surface}+\Delta I_{\rm surface} \, , 
\eea
where
\bea
 I^0_{\rm surface} &=& -\frac{\ell}{2 G}\left[ \frac{1}{2} \frac{1}{\delta^2}   + \Ordo(\delta)\right]\cr
 \Delta I_{\rm surface}(\gamma) &=&  -\frac{\ell}{2 G}\left[ \frac{2}{\delta} \, \alpha(\sqrt{1-2\gamma^2}) 
 + \Ordo(\delta)
 \right] \, . 
\eea
As a consistency check, we find in the limit $\gamma$ approaches zero that both $I^0_{\rm bulk}$ and $I^0_{\rm surface}$ agree with those without the Janus deformation. One also note that $\Delta I_{\rm bulk}(0)=\Delta I_{\rm surface}(0)=0$, as required. The details of the computation are again relegated to appendices \ref{appc} and \ref{appd}.

Thus, the renormalized free energy is found to be
\bea
I_{\rm ren}= F =F_{(0)} + F_I  \, , 
\label{2difreeold}
\eea
with
\bea
F_{(0)} &=& -\frac{\ell}{2 G}\log(\mu r) \cr
 F_I \ & =&  \frac{\ell}{4 G} \log {\sqrt{1-2\gamma^2}} \, . 
\label{2difree}
\eea
where $F_{(0)}$ is the renormalized free energy of undeformed CFT$_2$ and $F_I$   the interface free energy. Note that the latter
is independent of renormalization scheme as discussed in section \ref{sum}. 
We also recall that, for $\gamma \ll 1$, the massless minimal scalar field is expanded as
\bea
\phi (y) - \phi_0 = \gamma \tanh y +{1 \over 2} \gamma^3 \left( \tanh y + {1 \over 3} \tanh^3 y \right) + \cdots.
\eea
Interestingly, the total free energy is monotonically lowered by introducing the Janus interface of deformation $\gamma$.   
The renormalized partition function is given by $Z = Z_{(0)} \cdot  Z_I $,  where 
\bea
Z_{(0)} &=& (\mu r)^{\frac{c}{3}} \cr
Z_I  &=& \left[ \frac{1}{\sqrt{1-2\gamma^2} }\right]^{\frac{c}{6}} \, . 
\eea
We note that $F_I$ and hence $Z_I$ are independent of $r$. This reflects that the interface is an odd-dimensional conformal field theory, preserving the SO(2,1)
 conformal invariance. 
We can also relate, by a suitable conformal transformation, the free energy $-F_I$ to the interface entropy $S_I$, as was done in \cite{Azeyanagi:2007qj, Bak:2011ga, Jensen:2013lxa}. In turn, from the exponential of interface entropy, we also learn about the degeneracy of ground-states newly created by the presence of the interface. 

\section{Free Energy of Janus CFT$_3$}
For  arbitrary dimension $d+1$ of the gravity dual, we use the metric in the form
\bea
\rmd s^2 &=& \frac{\l^2}{q^2}\left[ \frac{\rmd q^2}{P(q)} + \frac{1}{\cos^2\lambda} (\rmd \lambda^2 + \sin^2 \lambda \rmd s^2 (\mathbb{S}^{d})) \right] \, . 
\eea
To get the $d$-dimensional boundary as $\mathbb{S}^d$, we choose the infrared cutoff hypersurface as
\bea
q(\lambda) &=& \frac{\delta_1}{\cos\lambda} \, , 
\eea
where $\delta_1$ is a dimensionless cutoff parameter introduced previously. Expressions for the bulk action and the surface action in arbitrary dimensions are complicated, so we relegate them in  appendix \ref{appe}. Rather, we specialize our computation to lowest even dimension, $(d+1)=4$. At zeroth order in $\gamma$, we find that  
\bea
I^{(0)}_{\rm bulk} &=& \frac{\l^2 \pi}{4G} \(\frac{1}{\delta_1^3} - \frac{3}{\delta_1} + 2 + \Ordo(\delta_1)\)\cr
&=& \frac{\l^2 \pi}{32 G} \(\frac{1}{\delta^3} - \frac{9}{\delta} + 16 + \Ordo(\delta)\)
\eea
and 
\bea
I^{(0)}_{\rm surface} &=& \frac{3\l^2\pi}{4G} \(-\frac{1}{\delta_1^3} + \frac{1}{\delta_1} +\Ordo(\delta_1)\)\cr
&=& \frac{3\l^2\pi}{32G} \(-\frac{1}{\delta^3} + \frac{1}{\delta} + \Ordo(\delta)\) \, . 
\eea
These are in perfect agreement with the previous computation based on the FG coordinate system.

For the corrections to first-order in $\gamma^2$, we find that
\bea
I^{(1)}_{\rm bulk} &=& \gamma^2 \frac{3\l^2 \pi}{32 G} \(\frac{1}{\delta_1^2} + \frac{2}{3} \log \delta_1 - 1 + \Ordo(\delta_1)\)\cr
&=&  \gamma^2 \frac{3\l^2 \pi}{32 G} \(\frac{1}{4\delta^2}-\frac{1}{2}+ \frac{2}{3} \log 2+ \frac{2}{3} \log \delta + \Ordo(\delta)\)
\eea
and 
\bea
I^{(1)}_{\rm surface} &=& -\gamma^2 \frac{\l^2\pi}{64 G}\(\frac{13}{\delta_1^2} - 4+\Ordo(\delta_1)\)\cr
&=& -\gamma^2 \frac{\l^2\pi}{64 G}\(\frac{13}{4\delta^2} + \frac{5}{2} + \Ordo(\delta)\) \, . 
\eea
Summing up all the contributions, we get 
\bea
&& I_{\rm bulk}+I_{\rm surface} = \frac{\l^2 \pi}{4G} \(-\frac{2}{\delta_1^3} + 2\) + \gamma^2\frac{\l^2 \pi}{64 G} \(-\frac{7}{\delta_1^2} - 2 + 4 \log \delta_1 \) + \Ordo(\gamma^4)\cr
&&   = \frac{\l^2 \pi}{4G} \(\ppp-\ppp
\frac{1}{4\delta^3}\ppp-\ppp \frac{3}{4\delta}+2\)+\gamma^2 \frac{\l^2 \pi}{64 G}\(\ppp-\ppp\frac{7}{4\delta^2}+ 4 \log \delta -\frac{11}{2}+ 4\log {2}\) \ppp + \Ordo(\gamma^4) \, . 
\eea
Thus the renormalized free energy is given by
\bea
I_{\rm ren}=F= F_{(0)} + F_I
\, ,
\label{3difreeold}
\eea
with
\bea
F_{(0)} &= &\frac{\pi \l^2 }{2G} \cr
  F_I \ &=& - \gamma^2 \frac{\l^2 \pi}{16 G} \log (\mu r)   + \Ordo(\gamma^4)
\label{3difree}
\eea
whose detailed identification is discussed in section \ref{sum}.
We see again that the interface contribution to the free energy shows the structure of CFT$_2$. 
In particular, being described by even-dimensional conformal field theory, the interface free energy exhibits $\log r$ dependence, indicating the Weyl anomaly of the interface. \footnote{See \cite{weyl} for  related studies of anomaly in the presence of boundaries.}

\section{Stress Tensor One-Point Function}
To further probe the interface, we study another physical observable, the one-point function of the stress tensor in CFT$_d$ and its change under the Janus deformation. We will again extract this observable from the gravity dual and compare the result with the CFT$_d$. We shall begin with  $(d+1)=3$. As explained in the last section, the construction of FG coordinate system for the Janus geometry faces a difficulty in the wedge of the bulk region that emanates from the boundary interface. So we will first determine the stress tensor one-point function at an infinitesimal distance away from the interface. 

Note that, in the coordinate system in (\ref{d+2janus}), the cutoff hypersurface may be introduced by
\bea
q_\pm \cos \lambda = \frac{\epsilon_1}{r}
\eea   
This generalizes the $d=2$ case of the previous section.  We then consider the region of the surface specified by
\bea
\cos \lambda \ \ge \ \frac{\tilde\epsilon}{r}
\eea
where we take $\tilde\epsilon$ infinitesimal but with $\tilde\epsilon/\epsilon_1 \gg 1$. 
This condition implies that the region of interest is at least an infinitesimal distance away from the interface from the viewpoint of the boundary space. In this region of the cutoff hypersurface, $q_\pm$ becomes infinitesimal:
\bea
q_\pm =\frac{\epsilon_1}{r \cos \lambda} \ \le \ \frac{\epsilon_1}{\tilde\epsilon} \ \ll \ 1 \, . 
\eea
Then, in the metric of (\ref{d+2janus}),  the function $P(q_\pm)$ can be replaced by $1-q^2_\pm$ 
while the terms ignored are of sufficiently higher order that they can be dropped off when evaluating the holographic stress tensor.
With such replacement, the metric becomes the undeformed one.
Now, in the FG coordinate system of the metric given in (\ref{fgmetric}), the metric $h_{ij}(x,u)$ is expandable in general in the form
\bea
h_{ij}= h^{(0)}_{ij} + u^2  h^{(2)}_{ij} + \cdots + u^d   h^{(d)}_{ij} + u^d \log u^2   \tilde{h}^{(d)}_{ij} +\cdots
\eea
where, for the undeformed case, the logarithmic term is present only when $d$ is even.
Here, $ h^{(0)}_{ij}$ is the metric for the boundary space which will also be denoted by $h^{B}_{ij}$.

For $d=2$, the one-point function of the boundary stress tensor is given by \cite{Skenderis:2002wp}
\bea
\langle T_{ij} \rangle = \frac{\ell}{8 \pi G} \left[ 
h^{(2)}_{ij} -h^{(0)}_{ij} h^{(2)}_{kl} h^{(0)\, kl} 
\right] +\tau_{ij} \, , 
\eea
where $\tau_{ij}$ is the contribution of the minimal scalar field to the stress tensor 
\bea
 \tau_{ij} = \frac{\ell}{8 \pi G} \left[ 
\partial_{i} \phi_B \partial_{j} \phi_B  -\frac{1}{2}h^{(0)}_{ij} h^{(0)\, kl} \partial_{k} \phi_B \partial_{l} \phi_B 
\right] \, . 
\eea 
Here, we denote by $\phi_B({\boldsymbol \Omega})$ the boundary value of the minimal scalar field $\phi$. As we are away from the interface, gradients of $\phi_B$ vanishes and hence $\tau_{ij}=0$ \footnote{Along the interface, this expression becomes singular and one needs some other method to fix it (see below).}. Using the metric in 
(\ref{dualsphere}), one finds for $\theta \ne {\pi \over 2}$ that 
\bea
\langle T_{ij} \rangle = \frac{c}{24\pi r^2} \, h^{B}_{ij} \, , 
\eea 
where $h_{ij}^B$ is the boundary value of metric field $h_{ij}$. This expression coincides with that of the undeformed case. We shall fix possible contribution at the 
interface location from our expression of the free energy.  Since the above expression is independent of our deformation 
parameter, one may compare the known result of the CFT on a sphere. One has in general
\bea
\langle T^i_{\phantom{i}\,i} \rangle_{\rm CFT} = \frac{c}{24 \pi} R(h_{B}) = \frac{c}{12\pi r^2} \, , 
\label{anomaly} 
\eea 
which follows from the Weyl anomaly of the CFT$_2$.  By the Lorentz invariance,  
this implies that
\bea
\langle T_{ij} \rangle_{\rm CFT} = \frac{c}{24\pi r^2} \, h^{B}_{ij} \, , 
\eea
which agrees with our computation above.

One can also check the stress tensor one-point function from our expression of the renormalized free energy. Varying the free energy with respect to $r$, one gets
\bea
\delta I^{(0)}_{\rm ren} = -\frac{1}{2} \int \rmd^2 q \, \sqrt{h_B} \, \, \delta h_{B}^{ij} \langle T_{ij} \rangle_{\rm CFT}  =- \frac{c}{3}\, \delta \log r \, . 
\eea 
This is consistent with the trace of the stress tensor in (\ref{anomaly}). On the other hand, 
as $\Delta I_{\rm ren} $ is independent of $r$, its variation with respect to $r$ implies that the interface contribution
$\langle  \Delta T^i_{\phantom{i}\,i} \rangle $ vanishes for $d=2$, which further implies that
\bea
\langle  \Delta T_{ij} \rangle =0 \, . 
\eea

For $d=3$, it is straightforward to show that  the stress tensor one-point function vanishes away from the interface location, which agrees with the stress tensor of the undeformed case. In fact, one can show that the undeformed holographic stress tensor from (\ref{dualsphere}) vanishes for any odd-dimensional sphere.
On the other hand, for $d=3$, the interface contribution is non-vanishing, 
\bea 
F_{I} = - \frac{c_{\rm eff}(\gamma)}{3} \log  \mu r  \, , 
\eea 
where, to leading order in $\gamma$, we read off from (\ref{3difree})
\bea
 c_{\rm eff}(\gamma)  = \ \frac{3\pi \ell^2}{16 G} \gamma^2 +\Ordo(\gamma^4)\,.
\eea
From these expressions together with the unbroken $SO(3)$ symmetry of our Janus solution, 
one obtains that $\langle  \Delta T_{ij} \rangle$ is given by
\bea
&& \langle  \Delta T_{\theta\theta} \rangle =\langle  \Delta T_{\theta\alpha} \rangle =0 
\cr
&& \langle   \Delta T_{\alpha\beta} \rangle = \frac{c_{\rm eff}}{24\pi r^2} \, h^{B}_{\alpha\beta} \, \delta\Big(\theta -\frac{\pi}{2}\Big) \, , 
\eea 
where $\alpha, \beta$ denote the directions along the interface. 
This again demonstrates that the interface contribution to the stress tensor is consistent with the expected structure for CFT on a sphere of one lower dimensions.

For completeness, we also record here the expression for the one-point function of the exactly marginal scalar operator:
\bea
\langle  {\cal O}_\phi \rangle_{\rm ICFT} = \frac{\ell^{d-1}}{8\pi G}\, \, \frac{\epsilon \Big(\frac{\pi}{2}-\theta \Big)  }{r^d \, |\cos \theta|^d}\, \gamma \, , 
\eea   
which we obtained from the gravity dual by the rules of AdS/CFT correspondence.

\section{Conformal Perturbation Theory}
So far, we computed one-point functions of ICFT$_{d-1}$ from the gravity dual. In this section, we compute them directly from the dual CFT$_d$ in the regime the deformation is weak. In this regime, we can use the conformal perturbation theory. Here again, we focus on $d=2$, but the method is applicable straightforwardly in arbitrary dimensions. 

Consider a CFT$_d$ perturbed by a local operator ${\cal O}_\phi$. The Lagrangian density of the theory is given by 
\bea
{\cal L} ({\boldsymbol \Omega}) = {\cal L}_0 ({\boldsymbol \Omega}) +\phi_B(\boldsymbol \Omega)\,{\cal O}_\phi ({\boldsymbol \Omega}) \, , 
\eea  
where the deformation coupling parameter (\ref{couplingparameter}) is expanded to the leading order in $\phi_B$.
We shall compute the one-point functions perturbatively in terms of the correlation functions of undeformed CFT$_d$.  
We assume that $\langle {\cal O}_\phi \rangle_{\rm CFT}=0$. Then, the leading-order correction to the free energy is identifiable as second order effect of the deformation 
\bea
\Delta F &=& -\frac{1}{2!} \int \hskip-0.3cm \int \rmd^2 {\boldsymbol \Omega} \sqrt{h_{B}(\boldsymbol \Omega) }\, \phi_B(\boldsymbol \Omega) \, \rmd^2 {\boldsymbol \Omega'} \sqrt{h_{B}(\boldsymbol \Omega')}\, 
\phi_B(\boldsymbol \Omega') \, \langle{\cal O}_\phi(\boldsymbol \Omega)
{\cal O}_\phi(\boldsymbol \Omega')
\rangle_{\rm CFT} \cr
&+& \cdots
\eea 
Here, $h^B_{ij}$ denotes the metric of the boundary space, and the ellipse denotes higher-order correction $O(\phi_B^3)$. Taking advantage of the conformal invariance, we can map the computation to $\mathbb{R}^2$. We start from $\mathbb{S}^2$ and describe it in terms of $\mathbb{R}^2$ variables ${\bf q} = (q_1, q_2)$ via stereographic projection
\bea
\rmd s_B^2=  \frac{4 r^2}{(1+ {\bf q}^2)^2} \rmd {\bf q}^2  
\, . 
\eea
Under the Weyl transformation
\bea
h^B_{ij}  &\rightarrow &  \ \Phi^2 ({\bf q}) \, h^B_{ij}  \, , 
\eea
the exactly marginal scalar operator with $\Delta  =2$ transforms as
\bea
{\cal O}_{\phi} &\rightarrow &   {1 \over \Phi^2 ({\bf q})} {\cal O}_\phi \, . 
\eea
With the choice of 
\bea
\Phi({\bf q}) = \frac{1+{\bf q}^2 }{2r} \, , 
\eea
we are mapping the two-sphere $\mathbb{S}^2$ to a two-plane $\mathbb{R}^2$ charted by the Cartesian coordinates ${\bf q}$. The equator of $\mathbb{S}^2$
is conformally mapped to a circle of unit radius on $\mathbb{R}^2$
\bea
{\bf q}^2  =1 \, . 
\eea 
On $\mathbb{R}^2$, the two-point function of the exactly marginal scalar operator ${\cal O}$ is given by
\bea
 \langle {\cal O}_\phi(\bq)
{\cal O}_\phi(\bq') \rangle_{\rm CFT} =\frac{{\cal N}}{\left[ ({\bf q}-{\bf q'})^2 + 2\kappa^2\right]^2 } 
\label{twopoint}
\eea
where we introduce a ultraviolet regulator $2\kappa^2$. As the scalar field dual to the operator ${\cal O}_\phi$ is normalized in the gravity side as in (\ref{einstein}), the normalization factor in (\ref{twopoint}) is fixed as
\bea
{\cal N} =\frac{\ell}{4 \pi^2 G}
\eea 
by the standard dictionary of the AdS/CFT correspondence in \cite{Freedman:1998tz}.
We then further perform a coordinate transformation the two-plane
from $\mathbb{R}^2$ to a cylinder $\mathbb{R} \times \mathbb{S}^1$ defined by
\bea
q_1 + i q_2 = \exp ({\sigma_1 + i \sigma_2}) \, . 
\eea
Here, $-\infty < \sigma_1 < \infty$ and $0 \le  \sigma_2 \le  2\pi $. The equator in $\mathbb{S}^2$, where the Janus interface is to be placed, is mapped in this coordinates to $\sigma_1=0$.  
Thus, with
$\phi_B = \phi_I \, \epsilon\( 
\frac{\pi}{2} -\theta
\)$, the interface contribution to the free energy becomes
\bea
 \Delta F\ppp = \ppp-\ppp\frac{\phi^2_I }{8} \ppp\ppp \int^{\frac{L}{2}}_{\ppp-\ppp \frac{L}{2}}\ppp\ppp\ppp \rmd\sigma_1\ppp \ppp 
\int_0^{2\pi}\ppp\ppp\ppp \ppp\ppp d\sigma_2 \ppp\ppp
\int^{\frac{L}{2}}_{\ppp-\ppp\frac{L}{2}}\ppp \ppp\ppp  \rmd \sigma'_1 \ppp\ppp \int_0^{2\pi}\ppp\ppp\ppp \ppp \ppp\ppp \rmd \sigma'_2 
\frac{  {\cal N} \, \epsilon (\sigma_1)\epsilon (\sigma'_1)}{\left[\cosh (\sigma_1 \ppp\ppp -\ppp\ppp\sigma'_1)  \ppp-\ppp\cos (\sigma_2\ppp\ppp-\ppp\ppp\sigma'_2)  \ppp+\ppp\kappa^2 \right]^2} \ppp+ \ppp O(\phi^3_I)        \, ,     
\eea
where we also introduce the infrared regulator $L$ by putting the system to a box of size $L$. This infrared regulator corresponds in the holographic gravity dual to an ultraviolet cutoff around 
the north and the south poles of the original $\mathbb{S}^2$.  This integral can be carried out  \cite{Bak:2007qw, Bak:2011ga}, and the result reads
\bea
\Delta F_{\rm reg} = -\frac{\ell }{2 G} \left(
\frac{L}{4\kappa^2} -\frac{1}{\sqrt{2}\kappa} + \frac{1}{2} \coth \frac{L}{2} + O(\kappa)
\right) \, \phi^2_I  +O(\phi^3_I)  \, . 
\eea
Thus, after renormalization, we get
\bea
F_I= -\frac{\ell }{4 G}\gamma^2  + O(\gamma^3)  
\eea
where we used the fact $\phi_I =\gamma +O(\gamma^2)$ for $d=2$ which can be identified from (\ref{icoeff}) and (\ref{scalarpro}).
We see that this agrees with our result (\ref{2difree}) in the gravity dual side, confirming the correspondence.

\subsection{Related Janus solutions}

\begin{figure}[t!]
\centering  
\includegraphics[width=11cm]{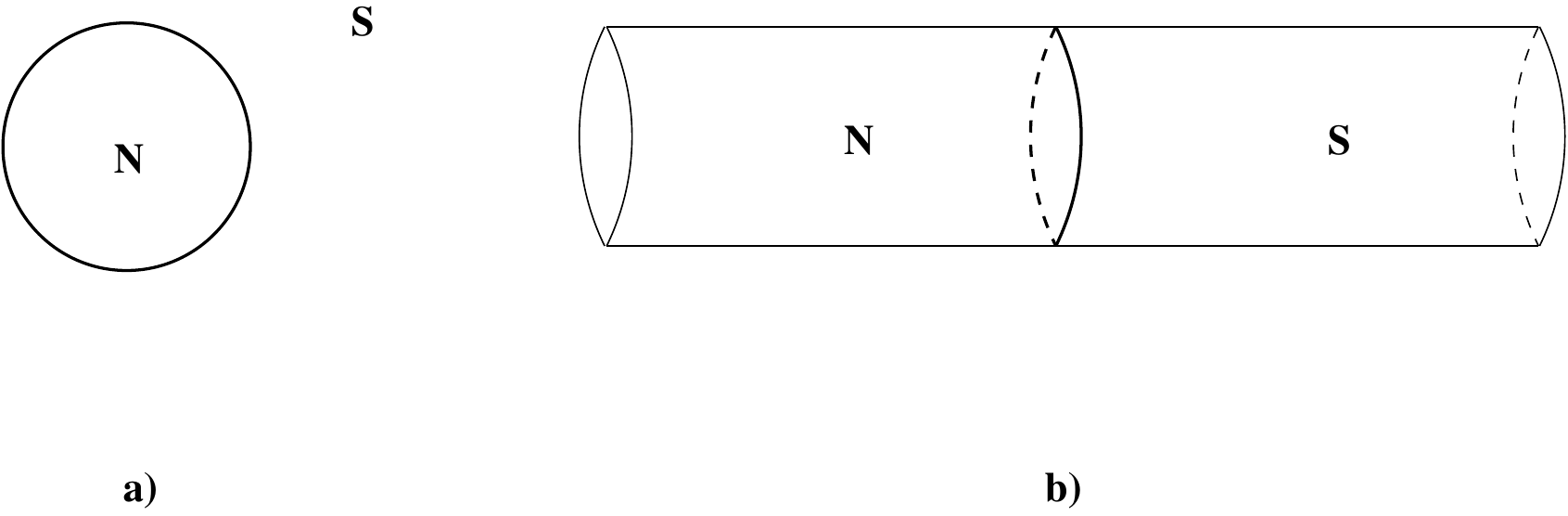}
\caption{\small  a) The Janus system on $\mathbb{R}^d$ with an interface of spherical shape is depicted. b) The Janus system on $\mb{R} \times \mathbb{S}^{d-1}$ with an interface at $\tau=0$
is depicted where $\tau$ is a coordinate along $\mb{R}$ direction.
On the $N$/$S$  region, we turn on an exactly marginal scalar operator with a coupling
$+\phi_I/-\phi_I$ respectively. 
}
\label{fig2}
\end{figure}

Motivated by the above  maps for the field theory side in this section,  we would like to obtain coordinates for the bulk Janus solution corresponding to the latter two ICFT's discussed above.  First we consider the ICFT on ${\mb{R}}^d$ with a spherical-shaped interface. The corresponding ICFT is depicted in Fig. \ref{fig2}a. To obtain this conformal boundary, we start by expressing the metric $ds^2_{M_d}$ in (\ref{d+2janus}) as
\bea
\rmd s^2_{M_d}=\frac{4}{(1-\xi^2)^2}\left( \rmd \xi^2 + \xi^2 \rmd s^2(\mathbb{S}^{d-1}) \, \right) \, , 
\eea
which is obtained by the coordinate transformation
\bea
\sin \lambda = \frac{2}{\xi +\xi^{-1}} \, . 
\eea
Here, $0\le \xi \le 1$ for the $q_+$ patch and $1\le \xi < \infty$ for 
the $q_-$ patch. At $q_\pm =q_*$, one needs an inversion coordinate transformation $\xi \rightarrow \frac{1}{\xi}$ to join 
the two patches in a smooth manner. To get the metric of the conformal boundary as a plane, we choose a defining function as
\bea
v(q_{\pm}) =   q_\pm \frac{|1-\xi^2|}{2 \ell}
\eea
which has a simple zero at the boundary $q_{\pm} = 0$. We then multiply the bulk metric by $v(q_{\pm})^2$ and take the limit $q_{\pm} = 0$. We then get the boundary metric as that of $\mb{R}^d$,
\bea
\rmd s^2_B =  \rmd \xi^2 + \xi^2 \, \rmd s^2(\mathbb{S}^{d-1}) \, . 
\eea
An alternative way to derive the bulk metric is to let $v$ be the new bulk coordinate replacing $q_{\pm}$. We then re-express the bulk metric in the new coordinate $v$ and find a double pole in the metric at $v=0$. We then multiply the bulk metric by the defining function $v^2$ and take the limit $v\rightarrow 0$. Again we will arrive at the same boundary metric. In these new boundary coordinates, the interface is located at the unit sphere $\mathbb{S}^{d-1} \subset \mb{R}^d$ at $\xi=1$. Of course, the radius of 
$\mathbb{S}^{d-1}$ can take any positive value by a rescaling of $v$ in the above. 

One can further make the coordinate transformation $\xi = e^{\tau}$ and get
\bea
\rmd s^2({\cal M}_d)=\frac{1}{\sinh^2 \tau }\left( \rmd \tau^2 +  \rmd s^2(\mathbb{S}^{d-1}) \, \right) \, . 
\eea
The $\tau$ coordinate is ranged over  $(-\infty, 0]$ for the $q_+$ patch while 
 $[0,  \infty)$ for the $q_-$ patch. If the boundary is identified with the coordinate $v=  q_\pm \sinh \tau$ with $v=0$,
the boundary space  becomes $\mb{R}\times \mathbb{S}^{d-1}$ with the metric
\bea 
\rmd s^2_B =\ell^2 \left( \rmd \tau^2 +  \rmd s^2(\mathbb{S}^{d-1}) \, \right) \, . 
\eea
The interface is located at $\tau=0$, as depicted in Fig. \ref{fig2}b.

\section{Interface Degrees of Freedom}
In a CFT on a torus, extra ground-state degeneracy is produced by the presence of a boundary (or a defect), whose number is
denoted by $g$. Then, $\ln g$ is identified with the `boundary entropy' counting degrees of freedom localized on the boundary.  
The $g$-theorem states that 
\bea
 \frac{\rmd}{\rmd l}\, g(l)  \le 0
\eea
with some length scale $l$, which is first suggested in Ref.~\cite{Affleck:1991tk}.  \footnote{See also the discussion on boundary/interface F-theorem in \cite{Gaiotto:2014gha}. }   In the situation the RG-flow is triggered by operators localized at the boundary, the $g$-theorem was proven in \cite{Friedan:2003yc}. However, as was shown in \cite{Green:2007wr} (see also explicit computation \cite{Bachas:2007td, Bachas:2012bj} in the free
field theory contexts), the $g$-function may either increase or decrease when the RG flow is triggered by operators of the bulk CFT. Here we examine interface counterparts of ground-state degenerarcy and g-theorem. On the two-sphere $\mathbb{S}^2$, the interface is localized along a circle, 
which may be interpreted as a circle of Euclidean time. Then $-\Delta F$ of the interface contribution can be interpreted as an interface entropy $S_I$.  Consider two interfaces  \cite{Bak:2013uaa} that are separated by a small distance $l$ around the equator. Let their interface coefficients are given by
$\phi_I$ and $\phi'_I$ which may have either signatures. In such circumstance, it is clear that
\bea
S_I (\phi_I,\phi'_I, l/r) \rightarrow  S_I (\phi_I+\phi'_I)
\eea
 as $ l/r \rightarrow 0$.  The entropy may either increase if ${\rm sign}\,\,\phi_I = {\rm sign}\,\, \phi'_I$ or decrease if 
 ${\rm sign}\,\,\phi_I =- {\rm sign}\,\, \phi'_I$. 
Therefore, as for the defect, we do not expect monotonicity property for the interface. 
Of course, this is expected since the perturbation induced by the interface is controlled by the bulk CFT operators. 
The same argument holds for higher-dimensional interfaces. We do not have any monotonicity of the interfaces degrees of freedom.

\section*{Acknowledgement}

We thank John Cardy, Yu Nakayama and Adam Schwimmer for useful discussions. We also acknowledge the ``Liouville, Integrability and Branes (11)"  Focus Program at the Asia-Pacific Center for Theoretical Physics for excellent collaboration environment. SJR acknowledges the NORDITA workshop ``Holography and Dualities 2016: New Advances in String and Gauge Theory", where this work was completed. DB was supported in part by the National Research Foundation grant 2014R1A1A2053737. SJR was supported in part by the National Research Foundation of Korea grants 2005-0093843, 2010-220-C00003 and 2012K2A1A9055280.


\appendix
\section{Coordinates on AdS$_{d+1}$}

The Euclidean AdS$_{d+1}$ is described  in $\mathbb{R}^{1,d+1}$ of Cartesian coordinates $(x^0, {\bf x})$
by the hypersurface 
\bea
X^M X^N \eta_{MN} = 
- (X^0)^2 + {\bf X}^2 = - \ell^2 .
\eea
We want to  introduce coordinates of this space. Depending on how we foliate the space, there are three independent coordinate systems.

\subsection{Global, AdS slice, Poincare Patches}
We may slice the Euclidean AdS$_{d+1}$ by foliations of $\mathbb{S}^d$.  This leads to global patch coordinates, given by
\bea
\begin{matrix} X^0 &= & \l \cosh \rho &\in \quad [1, \infty) \\
{\bf X} &= & \hat {\bf e} \, \l \sinh \rho  &\in \quad \mathbb{R}^{d+1} \, .
\end{matrix}
\eea
Here, $\rho \in [0, \infty)$ and $\hat {\bf e}$ is a Euclidean vector on $\mathbb{S}^d$ of unit radius. The $\mathbb{S}^d$ may be further parametrized by $\hat {\bf e} = ({\bf e} \sin\theta,\cos\theta)$,  where ${\bf e}$ is a vector on $\mathbb{S}^{d-1}$ of unit radius. 
We can alternatively introduce a compact coordinate $\lambda \in [0,\pi/2]$ such that $\sinh \rho = \tan \lambda$. In terms of either variable, the Euclidean AdS$_{d+1}$ metric is given by
\bea
\rmd s^2 &=& \l^2 \(\rmd \rho^2 + \sinh^2 \rho \, \rmd s^2(\mathbb{S}^d) \, \)\cr
\cr
&=& \frac{\l^2}{\cos^2\lambda} \(\rmd \lambda^2 + \sin^2 \lambda \,  \rmd s^2(\mathbb{S}^d) \, \) \, . 
\eea

We may also slice the Euclidean AdS$_{d+1}$ by foliations of AdS$_d$. The leads to AdS slicing coordinates, given by
\bea
\begin{matrix}
X^{\mu} & = & \hat{ \mathfrak{n}} \, \l \cosh y & \in & [1, \infty) \times \mathbb{R}^d  \\
X^{d+1} & = & \l \sinh y & \in & \mathbb{R}
\end{matrix} \, . 
\eea
Here,  $y \in [-\infty,+\infty]$ and $\hat{\mathfrak{n}}$ is a Lorentzian vector on AdS$_d$ of unit radius. 
The AdS$_d$ may be further parametrized by $\hat{\mathfrak{n}}  = (\sec\lambda, \mathfrak{n} \tan\lambda )$, where $\mathfrak{n}$ is a vector on AdS$_{d-1}$  of unit radius. The metric is given in terms of these coordinates by 
\bea
\rmd s^2 &=& \l^2 \( \rmd y^2 + \cosh^2 y \, \rmd s^2({\rm AdS}_d) \, \) \, . 
\eea

The above two coordinate systems are related by the diffeomorphism
\bea
\cosh \rho &=& {\cosh y} \sec \lambda\cr
\sinh \rho \sin \theta &=& \cosh y \tan \lambda\cr
\sinh \rho \cos \theta &=& \sinh y
\label{diffeo}
\eea
Thus,  we see that $\theta\in[0, \frac{\pi}{2}]$ and $\theta\in[\frac{\pi}{2},\pi]$ intervals in the global coordinate is mapped to $y\in[-\infty,0]$ and $y\in[0,\infty]$ intervals in the AdS slice coordinate, respectively. 

Finally, we may also slice the Euclidean AdS$_{d+1}$ by foliation of $\mathbb{R}^d$. This leads to the Poincare patch coordinates given by
\bea
\begin{matrix}
X^0 &=& ({\l}/{2}) \[\, z \, + \, \(1+x^i x^i\)/z \, \] &\in & \mathbb{R}^+  \\
{\bf X} &=& \l ({\bf x}/{z}) & \in & \mathbb{R}^d  \\
X^{d+1} &=&({\l}/{2})\[ \, z+\ \(-1+x^i x^i\)/z \, \] &\in & \mathbb{R}
\end{matrix}
\eea
The metric is
\bea
\rmd s^2 &=& \frac{\l^2}{z^2} \(\, \rmd z^2 + \rm d {\bf x}^2 \, \) \, . 
\eea

\subsection{The conformal boundary}
We want to take the conformal boundary as a sphere. Clearly, this is most naturally described in the global coordinate. We want to know how this boundary looks like in the AdS slice coordinate. To do so, we define the boundary with the following infrared cutoff
\bea
e^{-\rho_{\infty}} = \delta \qquad \mbox{and} \qquad \frac{\cos\lambda}{\cosh y} = {\delta_1} \, . 
\eea
By the diffeomorphism (\ref{diffeo}), they are related as
\bea
\frac{1}{\delta_1} = {1 \over 2} \left( \delta+\frac{1}{\delta} \right)
\eea
and also as 
\bea
\left(\frac{1}{\delta}-\delta \right)\sin\theta &=& \left(\frac{1}{\delta} + \delta \right) \sin \lambda \, . 
\eea
On the other hand, using the relation
\bea
 \frac{\(\delta^{-1}-\delta\)^2}{\(\delta^{-1}+\delta\)^2} = 1 - \delta_1^2, 
 \eea
we see that the metric of cutoff $\mathbb{S}^d$ in the global coordinate 
\bea
\rmd s^2 &=& {\ell^2 \over 4} (\delta^{-1} - \delta)^2 \left( \rmd \theta^2 + \sin^2\theta \, \rmd s^2(\mathbb{S}^{d-1} ) \right) 
\eea
becomes in the AdS slice coordinate 
\bea
\rmd s^2 &=& \left( 1-\frac{\delta_1^2}{\cos^2\lambda} \right)^{-1} \rmd \lambda^2 + \frac{1}{1-\delta_1^2} \sin^2\lambda \, \rmd s^2(\mathbb{S}^{d-1}).
\eea

\section{Computation of $I_{\rm surface}$}\label{appb}
In terms of the coordinate 
\bea
v = r \frac{\cos \lambda}{\cosh y} \, , 
\label{surface}
\eea
the boundary surface is specified by the hypersurface $v = \epsilon_1$, where $\epsilon_1$ is a cutoff. On the surface, the $y$ coordinate ranges over $ [-y_{\infty},  y_\infty]$, where $\cosh y_{\infty} = \frac{r}{\epsilon_1}$. We also find that $
\cos \lambda \in  \[\frac{\epsilon_1}{r},1\]$ and we define $\cos  \lambda_0=\frac{\epsilon_1}{r}$. 
We use ($\lambda,  \phi$) as  boundary surface coordinates. For a fixed $\lambda$, the coordinate $y$ has double roots, as seen from (\ref{surface}). Thus, the boundary surface can be covered by two branches of coordinates   $x^i = (\lambda, \phi )_+ \cup  ( \lambda, \phi )_- $ 
where $+/-$ refers to the part of surface with positive/negative $y$. 
We also denote bulk coordinates as $x^a = (v,x^i)$. Tangent vectors to the boundary are $\partial_i$. The normal vector is orthogonal to the tangent vectors with respect to the bulk metric $g_{ab}$. This means that
\bea
g_{ia} n^a = n_i  = 0 \, . 
\eea
The bulk metric can be expressed as
\bea
\rmd s^2 = N^2 dv^2 +  \gamma_{ij} (\rmd x^i + N^i \rmd v) (\rmd x^j + N^j \rmd v) \, . 
\label{lapsh}
\eea
By matching this with the metric of Euclidean $AdS_3$, we can identify shift and lapse functions $N, N^i$ as 
\bea
N &=& \frac{\ell}{v} \frac{1}{\sqrt{1-\frac{v^2}{r^2}}}\cr
N^{\lambda} &=& \frac{v}{r^2-v^2} \tan\lambda
\eea
Moreover, $\gamma_{ij}$ can be identified with the induced metric on the boundary. It has the non-vanishing components  
\bea
\gamma_{\lambda\lambda} &=& \frac{\ell^2}{1-\frac{v^2}{r^2\cos^2 \lambda}} \(\frac{r^2}{v^2} - 1\)\cr
\cr
\gamma_{\phi\phi} &=& \frac{\ell^2 r^2}{v^2} \sin^2\lambda \, . 
\eea
In these coordinates, the unit normal vector $n^a$ obeying $g_{ab} n^a n^b = 1$ is given by
\bea
n_a &=& -(N,0,0)\cr
n^a &=& -\frac{1}{N} (1,-N^i)
\label{normal}
\eea
We can then compute the extrinsic curvature 
\bea
K &=& \frac{1}{\sqrt{g}} \partial_a (\sqrt{g}\, n^a)
\eea
with the result
\bea
\sqrt{\gamma} K &=& \frac{2\ell r^2}{v^2} \frac{\sin\lambda}{\sqrt{1-\frac{v^2}{r^2\cos^2 \lambda}}} \, . 
\eea
We then obtain the integral  of the extrinsic curvature as
\bea
-\frac{1}{8\pi G} \, 2 \int_0^{\lambda_0} \rmd \lambda \int_0^{2\pi} \rmd \phi \sqrt{\gamma} \, K &=& - \frac{\ell }{G} \frac{r^2}{\epsilon_1^2} \sqrt{1-\frac{\epsilon_1^2}{r^2}} \, , 
\eea
where the extra factor $2$ in the left-hand side comes from the fact that we have two branches of boundary coordinates.

\section{Computation of $I_{\rm bulk}$ for ICFT$_2$}\label{appc}
With the cutoff surface defined in Section \ref{icft2}, the bulk integral takes the form
\bea
I_{\rm bulk}   = \frac{\ell}{G}\int ^{y_0}_0 \rmd y \,  f(y)
\int^{\lambda_y}_0 \frac{\rmd \lambda \, \sin \lambda }{\cos^2 \lambda} \, , 
\eea
where $y_0 \, (>0)$ is defined by the relation 
\bea
\sqrt{f(y_0)} = \frac{r}{\epsilon_1}
\eea
and $\cos \lambda_y = \frac{\epsilon_1}{r}\sqrt{f(y)}$. From this,  $e^{2 y_0}$ can be solved  in terms of $\epsilon_1$ by
\bea
e^{2 y_0}= \frac{4}{\sqrt{1-2 \gamma^2}} \left( \frac{r^2}{\epsilon^2_1}-\frac{1}{2}\right) + O(\epsilon^2_1) \, .
\eea 
For the regularization,  the $O(\epsilon_1^2)$ contribution is not needed.  Carrying out  the $\lambda$ integration, one gets
\bea
I_{\rm bulk}   = \frac{\ell}{G}\left[ \, \frac{r}{\epsilon_1}
\int ^{y_0}_0 \rmd y \, \sqrt{ f(y)}
-\int ^{y_0}_0 \rmd y \,  f(y)
 \right] \, . 
\label{integraly}
\eea
In this expression, the first integral can be rearranged as
\bea
\int ^{y_0}_0 \rmd y \, \sqrt{ f(y)} &=& \alpha(\sqrt{1-2\gamma^2})   +(1-2\gamma^2)^{\frac{1}{4}}\int ^{y_0}_0 \rmd y \,  \cosh y \cr
   &-&  \int ^\infty_{y_0} \rmd y \left[ \sqrt{ f(y)}
  -(1-2\gamma^2)^{\frac{1}{4}} \,  \cosh y 
\right] \, , 
\eea
where
\bea
\alpha(\sqrt{1-2\gamma^2}) \equiv   \int ^{\infty}_0 \rmd y \left[ \sqrt{ f(y)}
  -(1-2\gamma^2)^{\frac{1}{4}} \,  \cosh y  \right] \, . 
\eea
From this, it is straightforward to get
\bea
\frac{r}{\epsilon_1}\int ^{y_0}_0 \rmd y \, \sqrt{ f(y)} =
 \frac{r}{\epsilon_1} \alpha(\sqrt{1-2\gamma^2})  + \frac{r^2}{\epsilon^2_1}-\frac{1}{2}+ O(\epsilon^2_1) \, . 
\eea
Carrying out the second integral in (\ref{integraly}) explicitly, we get
\bea
I_{\rm bulk}&=&  \frac{\ell}{ G}\left[ \,\,
 \frac{1}{2}\left(\frac{r^2}{\epsilon^2_1}-\frac{1}{2}\right) +  \frac{r}{\epsilon_1} \alpha(\sqrt{1-2\gamma^2}) +\frac{1}{2}\log \frac{\epsilon_1}{2r} \right. 
\cr
 && \ \ \  \left. -\frac{1}{4}\log  \frac{1}{\sqrt{1-2\gamma^2}}
 + O(\epsilon^2_1)\,\, \right]
\eea
Finally, $\alpha(z)$ can be expressed in terms of the complete elliptic integrals and the relation (\ref{eps0eps1}) is used to rewrite
the above in terms of $\delta$ instead of $\epsilon_1$.

\section{Computation of $I_{\rm surface}$ for ICFT$_2$}\label{appd}
In this appendix, we work in the coordinates $(v,\lambda, \phi)_\pm$ introduced in section  \ref{icft2}. To simplify our presentation,
we shall introduce two quantities $D_d$ and $\tilde{D}_d$ , respectively, defined by
\bea
D_d (v,\lambda) &\equiv&   \sqrt{1- \left(\frac{v}{r \cos \lambda}\right)^{2} +\frac{\gamma^2}{d(d-1)}
\left(\frac{v}{r \cos \lambda}\right)^{2d}
} 
\cr
\tilde{D}_d (v,\lambda)  &\equiv&   \sqrt{1- \frac{v^2}{r^2 } +\frac{\gamma^2}{d(d-1)}\left(\frac{v}{r \cos \lambda}\right)^{2d}} \, . 
\eea 
The description here is in parallel with the treatment of appendix \ref{appb}, which is for the undeformed case. In the metric 
given in (\ref{lapsh}), the lapse and shift can be identified as
\bea
N &=& \frac{\ell}{v}\, \frac{1}{{\tilde{D}}^2_2}\cr
N^{\lambda} &=& \frac{v}{r^2} \frac{\tan\lambda}{{\tilde{D}}^2_2}
\eea
and the nonvanishing components of $\gamma_{ij}$ are given by
\bea
\gamma_{\lambda\lambda} &=& \frac{\ell^2r^2}{v^2} \frac{{\tilde{D}}^2_2 }{{{D}}^2_2}\cr
\gamma_{\phi\phi} &=& \frac{\ell^2 r^2}{v^2} \sin^2\lambda
\eea
Then, the unit normal vector is given by the form in (\ref{normal}). Thus, the extrinsic curvature contribution is 
identified as
\bea
\sqrt{\gamma} K &=& \frac{2\ell r^2}{v^2} {\sin\lambda} \left[
\frac{1}{D_2} +\frac{\gamma^2}{2} \left(\frac{v}{r \cos \lambda}\right)^{4} \frac{D_2}{{\tilde{D}}^2_2}
\right] \, . 
\eea
The integral over the boundary of the extrinsic curvature is given by
\bea
-\frac{1}{8\pi G} 2 \int_0^{\lambda_0} \rmd \lambda \int_0^{2\pi} \rmd \phi \sqrt{\gamma} K \, , 
\eea
where again we have an extra factor 2 and $\cos \lambda_0 = \frac{\epsilon_1}{r q_*}$.
Thus, by carrying out integral explicitly, one is led to 
\bea
I_{\rm surface}&=& - \frac{\ell }{G}\left[ \frac{r^2}{\epsilon_1^2}  -\frac{1}{2} +\frac{2r}{\epsilon_1} \alpha(\sqrt{1-2\gamma^2}) 
  + O(\epsilon^2_1)\, .
\right]
\eea
Again, this result can be written in terms of the cutoff $\delta$.

\section{General expressions for ICFT$_d$ to first order in $\gamma^2$}\label{appe}
\subsection{The bulk integral}
For a $d$-dimensional ICFT on $\mathbb{S}^d$, we have the following bulk metric 
\bea
\rmd s^2 &=& \frac{\l}{q^2}\(\frac{\rmd q^2}{P(q)} + \frac{\rmd \lambda^2 + \sin^2 \lambda \, \rmd s^2_{S^{d-1}}}{\cos^2\lambda}\) \, . 
\eea
Therefore, the bulk term is 
\bea
I_{\rm bulk} &=& 2\frac{d \l^{d-1}}{8\pi G} \int_{\mathbb{S}^{d-1}} d\Omega_{d-1} 
 \int_0^{\lambda_0} d\lambda \frac{\sin^{d-1} \lambda}{\cos^d\lambda} \int_{q(\lambda)}^{q_*} dq \frac{1}{q^{d+1} \sqrt{P(q)}}
\eea
Here, the metric function
\bea
P(q) &=& 1 - q^2 + \E q^{2d} \, , 
\eea
the expansion parameter
\bea
\E &=& \frac{\gamma^2}{d(d-1)}
\eea
related to the Janus deformation. The integration bounds are specified by
\bea
P(q_*) &=& 0\cr
q_* &=& \frac{\delta_1}{\cos\lambda_0}\cr
q(\lambda) &=& \frac{\delta_1}{\cos\lambda}
\eea
We first evaluate the last integral,
\bea
\Gamma(\E,q(\lambda)) &:=& \int_{q(\lambda)}^{q_*} \rmd q \frac{1}{q^{d+1} \sqrt{P(q)}}
\eea
to linear order in $\E$. We eliminate the $q^2$ term from $P(q)$ using
\bea
q_*^2 &=& 1 + \E q_*^{2d}
\eea
and change the integration variable as $q = q_* t$. We get 
\bea
P(q_* t) &=& (1-t^2) \(1 - \E q_*^{2d} t^2 \frac{t^{2(d-1)}-1}{t^2-1}\) \, . 
\eea
We then Taylor expand the integrand to first order in $\E$,
\bea
\Gamma(\E,q(\lambda)) &=& \frac{1}{q_*^d} \int_{\frac{q(\lambda)}{q_*}}^1 \rmd t \frac{1}{t^{d+1}\sqrt{1-t^2}} \(1 + \frac{1}{2} \E q_*^{2d} t^2  \frac{t^{2(d-1)}-1}{t^2-1}+\Ordo(\E^2)\) \, . 
\eea
Finally, we expand the prefactor $1/q_*^d$ and the lower boundary of integration in powers of $\E$, where the following expansion
\bea
q_* &=& 1 + \frac{1}{2}\E + \Ordo(\E^2)
\eea
for the smallest root is used. 
We find \cite{Bak:2015jxd}
\bea
\Gamma(\E,q(\lambda)) &=& \Gamma(0,q(\lambda)) + \frac{{}_2 F_1\(-\frac{1}{2},\frac{2-d}{2},2,1-q(\lambda)^2\)}{\sqrt{1-q(\lambda)^2}}\E + \Ordo(\E^2) \, 
\eea
where
\bea
\Gamma(0,q(\lambda)) &=& \int_{q(\lambda)}^1 \rmd t \frac{1}{t^{d+1} \sqrt{1-t^2}} \, . 
\eea
Next, we integrate over $\lambda$ to obtain the bulk term
\bea
I_{\rm bulk} &=& 2\frac{d\l^{d-1}}{8\pi G} V_{d-1} \int_0^{\lambda_0} \rmd \lambda \, \frac{\sin^{d-1}\lambda}{\cos^d\lambda} \, \Gamma(\E,q(\lambda)) \, . 
\eea
Here, we may expand the bound as $\lambda_0(\E) = \lambda_0(0) + \E \lambda'_0(0) + \Ordo(\E^2)$ and pick up a boundary term $\sim \E \lambda'_0(0) \Gamma(\E,q_*)$ but to linear order in $\E$ this is zero because $\Gamma(0,1) = 0$. We get
\bea
I_{\rm bulk} &=& \frac{d\l^{d-1}}{4\pi G} V_{d-1} \int_0^{\lambda_0(0)} \rmd \lambda \frac{\sin^{d-1}\lambda}{\cos^d\lambda} \Gamma(\E,q(\lambda)) \, , 
\eea
where $\cos\lambda_0(0) = \delta_1$. Changing variable of integration to $x=\cos\lambda$, we finally get
\bea
I_{\rm bulk} &=&  \frac{d\l^{d-1}}{4\pi G} V_{d-1} \int_{\delta}^1 \rmd x \frac{(1-x^2)^{\frac{d-2}{2}}}{x^d} \Gamma\(\E,\frac{\delta_1}{x}\) \, . 
\eea

\subsection{The surface term}
Using the same technique as we used for the case $d=2$, we find
\bea
\sqrt{\gamma}\, K &=& \l^{d-1} d \(\frac{r}{v}\)^d \sin^{d-1} \lambda \(\frac{1}{D_d} +
 \E \(\frac{v}{r\cos\lambda}\)^{2d} \frac{D_d}{\t D_d^2}\) \, . 
\eea

\end{document}